\documentclass[12pt,letterpaper]{article}
\usepackage{amssymb}
\usepackage{amsmath}
\usepackage{amsfonts}
\usepackage{epsfig}
\usepackage{feynmf}
\usepackage{color}
\newcommand{\sect}[1]{\setcounter{equation}{0}\section{#1}}

\usepackage{bm}
\outer\def\beginsection#1\par{\medbreak\bigskip
      \message{#1}\leftline{\bf#1}\nobreak\medskip
\vskip-\parskip
      \noindent}

\newcommand{\eq}{\begin{equation}}
\newcommand{\eqx}{\end{equation}}
\newcommand{\eqn}{\begin{eqnarray}}
\newcommand{\eqnx}{\end{eqnarray}}
\newcommand{\bi}{\begin{itemize}}
\newcommand{\ei}{\end{itemize}}

\def\e{\epsilon}

\def\z{\zeta}
\def\m{\mu}
\def\n{\nu}
\def\r{\rho}
\def\s{\sigma}

\def\t{\tau}
\def\p{\partial}

\def\a{\alpha}
\def\b{\beta}
\def\d{\delta}
\def\l{\lambda}
\def\g{\gamma}

\def\w{\omega}
\def\W{\Omega}

\def\nb{\nonumber}
\def\ap{\a {\rm '}}

\def\be{\begin{equation}}
\def\ee{\end{equation}}
\def\ba{\begin{eqnarray}}
\def\ea{\end{eqnarray}}

\begin{document}
\begin{titlepage}
\hfill \hbox{CERN-PH-TH/2010-190}
\vskip 0.1cm
\hfill \hbox{LPTENS-10/30}
\vskip 0.1cm
\hfill \hbox{NORDITA-2010-55}
\vskip 0.1cm
\hfill \hbox{QMUL-PH-10-09}
\vskip 1.5cm
\begin{flushright}
\end{flushright}
\vskip 1.0cm
\begin{center}
{\Large \bf  High-energy string-brane scattering: \\
leading eikonal and beyond} \vskip 1.0cm {\large Giuseppe
D'Appollonio$^{a, b}$, Paolo Di Vecchia$^{c, d}$,
Rodolfo Russo$^{e}$, \\
Gabriele Veneziano$^{f, g}$ } \\[0.7cm]
{\it $^a$ Dipartimento di Fisica, Universit\`a di Cagliari and
INFN\\ Cittadella
Universitaria, 09042 Monserrato, Italy}\\
{\it $^b$ Laboratoire de Physique Th\'eorique de L'Ecole Normale Sup\'erieure\\
24 rue Lhomond, 75231 Paris cedex, France}\\
{\it $^c$ The Niels Bohr Institute, Blegdamsvej 17, DK-2100 Copenhagen, Denmark}\\
{\it $^d$ Nordita, Roslagstullsbacken 23, SE-10691 Stockholm, Sweden}\\
{\it $^e$ Queen Mary University of London, Mile End Road, E1 4NS London, United Kingdom}\\
{\it $^f$ Coll\`ege de France, 11 place M. Berthelot, 75005 Paris, France}\\
{\it $^g$Theory Division, CERN, CH-1211 Geneva 23, Switzerland}
\end{center}
\begin{abstract}
We extend previous techniques for calculations of transplanckian-energy
string-string collisions to the high-energy scattering of
massless closed strings from a stack of $N$ Dp-branes in Minkowski
spacetime.
We show that an effective non-trivial metric emerges from the string
scattering amplitudes by comparing them against the
semiclassical dynamics of high-energy strings in the extremal p-brane
background.
By changing the energy, impact parameter and effective
open string coupling $\lambda = g N$, we are able to explore various
interesting regimes and to reproduce classical expectations,
including tidal-force excitations, even beyond the leading-eikonal
approximation.
\end{abstract}

\end{titlepage}

\sect{Introduction}
\label{intro}

One of the most remarkable properties of string theory is the generic
presence, in its massless spectrum, of a spin-2 particle behaving as
the graviton of general relativity. Invariance under general
coordinate transformations emerges as part of the larger gauge
symmetry of string theory and string interactions are automatically
compatible with it. This is clearly reflected in the low-energy string
effective action, which can be obtained from the scattering amplitudes
of the massless excitations and which reduces to an extended -- and
typically supersymmetric -- version of general relativity with
additional short-distance corrections. As a result, in string theory
spacetime is dynamical, even if the perturbative formulation of the
theory requires the choice of a fixed background. Any highly massive
or energetic state should then produce physical effects that can be
ascribed to an {\it effective} spacetime metric.

In a series of papers~\cite{Amati:1987wq, Amati:1988tn,
Amati:1990xe}\footnote{See also~\cite{Sundborg:1988tb}
for a relevant string-loop calculation and~\cite{'tHooft:1987rb},
\cite{Muzinich:1987in} for other derivations of the leading eikonal in
the field theory limit. A different approach~\cite{Gross:1987kza} is
valid at fixed (and even large) angle within a
high-energy window. The two methods agree in the intersection of their
respective regions of validity.} Amati, Ciafaloni and one of us
(referred to hereafter as ACV) analysed the way in which classical
gravity effects emerge from string computations performed in flat
Minkowski space and developed methods to derive quantum and string
corrections to general relativity. The process considered in ACV is
the collision of two strings at trans-planckian energies in a flat
Minkowski background. Given the strong gravitational fields generated
in the process, one expects that the effective spacetime experienced
by the two colliding quanta will be modified in a drastic way,
possibly leading to gravitational collapse and black hole formation.

In~\cite{Amati:1987wq} it was shown that an eikonal resummation of
the leading high-energy contributions to the S-matrix leads to
results that are fully consistent with general relativity
expectations whenever effects due to the finite string size ($l_s =
\sqrt{\alpha' \hbar}$) can be neglected. In this particular process
the effective geometry turns out to be, to leading order, the
well-known Aichelburg-Sexl (AS) shock-wave
metric~\cite{Aichelburg:1970dh} produced by an energetic pointlike
massless particle. This simple and intuitive interpretation
unfortunately fails beyond the leading-eikonal (or small deflection
angle) approximation~\cite{Amati:1990xe}. In general the effective
geometry produced by the collision is expected to be very
complicated and to depend on the system as a whole rather than on
each individual colliding string. An all-order approach was proposed
in \cite{Fabbrichesi:1993kd} (see also~\cite{Veneziano:1994bb}) but
it turned out to be too difficult to implement this beyond the
leading-eikonal approximation.

Recently there has been some
progress~\cite{Amati:2007ak,Marchesini:2008yh,Veneziano:2008zb,
  Veneziano:2008xa,Ciafaloni:2008dg} in trying to compute these
``classical corrections" to all orders through a dimensionally-reduced
effective action, but the approximations made are not yet fully under
control. Nonetheless these approximations seem to lead to physically
sensible results, at least in terms of identifying semiquantitatively
the critical parameters for gravitational collapse and of comparing
them with those based on the formation of a trapped surface
\cite{Eardley:2002re}.

Several effects due to a non-vanishing string length $l_s \ne 0$ were
also computed in~\cite{Amati:1987wq, Amati:1988tn,
Veneziano:2004er}: some of them do have a general relativity
interpretation (e.g.  as the effect of tidal forces exerted by an AS
metric on extended objects~\cite{Giddings:2006vu}) while others (like
the possible absorption of the elastic channel due to s-channel
formation of heavy strings) do not.  On the whole, a picture emerges
whereby string-size effects prevent gravitational collapse when the
Schwarzschild radius of the would-be back hole is smaller than the
string length parameter $l_s$ while the approach to gravitational
collapse is characterized, at the quantum level, by a rapid increase
in multiplicity and by the corresponding softening of the final quanta
\cite{ Veneziano:2004er, Amati:2007ak,Veneziano:2008xa}. The
transition to the black-hole formation regime, which resembles a phase
transition in general relativity, may turn out to be smoother in the
quantum case.

In this paper we shall apply the approach developed by ACV to the
study of a different process, the scattering of a closed string from
a stack of $N$ parallel Dp-branes in Minkowski spacetime. The
D-branes are massive solitons for which a microscopic string
description is available~\cite{Polchinski:1995mt}. This important
property makes the string-brane system an ideal framework to
understand the way in which string scattering amplitudes evaluated
in flat space can provide information about the dynamics in an
effective curved spacetime\footnote{There is an analogue of this in
quantum field theory: as shown long ago by Duff \cite{Duff:1974xx},
a class of tree diagrams for the scattering of a test particle from
a classical source reproduces the physical effects of the effective
Schwarzschild metric generated by the source. The difference is that
in string theory we have a microscopic quantum description of the
source itself and of its couplings to the test particles.}. Indeed,
from the point of view of perturbative string theory the presence of
a collection of Dp-branes is entirely taken into account by the
addition of an open string sector with suitable boundary conditions
and does not require any modification of the background. On the
other hand, from the point of view of the low-energy effective field
theory the Dp-branes are a massive charged state and their presence
will necessarily result in a curved spacetime.

The backreaction of the D-brane system
on spacetime is expected to be well-described
by the extremal p-branes
\cite{Horowitz:1991cd}, which are BPS solutions of the supergravity
equations of motion with a non-trivial metric, dilaton and
Ramond-Ramond $(p+1)$-form potential. For $p < 7$ and in the string
frame the extremal p-brane solution is given by
\begin{eqnarray}
&& d {{s}}^2 = \frac{1}{\sqrt{H(r)}}
\left(\eta_{\alpha\beta}dx^\alpha dx^\beta \right) + \sqrt{H(r)}
(\delta_{ij}d x^idx^j) \ ,
\label{metri}  \\
&&{\rm e}^{{{\phi}}(x)} = g\left[ H(r) \right]^{\frac{3-p}{4}} \ ,
\hspace{1cm} {\cal{C}}_{01...p}(x) =  \frac{1}{H(r)} - 1 \ ,
\label{dil}
\end{eqnarray}
where the indices $\alpha,\beta,\ldots$ run along the Dp-brane
world-volume, the indices $i,j,\ldots$ indicate the transverse
directions and $ r^2 = \delta_{ij} x^i x^j$. Finally
\begin{eqnarray}
 H (r) = 1 + \left( \frac{R_p}{r}\right)^{7-p} \ , \hspace{1cm}
R^{7 -p}_{p}  = \frac{g N (2 \pi \sqrt{\alpha'})^{7-p}}{(7-p)
\Omega_{8-p}} \ , \hspace{1cm} \Omega_n = \frac{2
\pi^{\frac{n+1}{2}}}{\Gamma ( \frac{n+1}{2})} \ ,
\label{eqparam}
\end{eqnarray}
where $g$ is the dimensionless string coupling constant, $N$ the
number of Dp-branes and $\W_n$ the volume of the $n$-dimensional unit
sphere. This effective description should be reliable as long as the
curvature is small in string units.  Evidence that $N$ parallel
Dp-branes correspond to the curved spacetime given in
Eqs.~(\ref{metri}) and~(\ref{dil}) was provided
in~\cite{Garousi:1996ad, Frau:1997,Bertolini:2000jy} where it was
shown that the large distance behaviour of the classical solutions can
be recovered from string-brane scattering amplitudes at tree level.

We shall analyse the system of $N$ parallel Dp-branes using as a
probe a high-energy massless closed string.  In the high-energy
limit the dominant interaction will be gravity and we will not be
sensitive to the other background fields. We shall also work in the
regime of small string coupling ($g \ll 1$) and we shall therefore
disregard all non-planar diagrams since their contribution is
suppressed by powers of $g$.  On the other hand, if we keep $\lambda
= g N$ fixed (and thus let $N \to \infty$) contributions with many
open string loops (boundaries) are not suppressed.  This is a
welcome feature: indeed, in the regime where the energy $E$ of the
closed string is very large the scattering amplitude grows with
energy in such a way that partial-wave unitarity breaks down at any
finite order.  These large contributions to the S-matrix from
diagrams with a different number of boundaries are expected to
combine together and yield a unitary eikonal form for the resummed
amplitude.

As in the case of the string-string scattering, our ultimate aim is
to calculate the eikonal phase to all orders in the classical and
string corrections. It will turn out that the problem of string
scattering from a stack of D-branes is much easier to deal with than
the original ACV-type collision problems. Indeed, the former problem
is similar to that of scattering in an external field with the
closed string acting as a probe and not back-reacting on the
geometry. As a result, while the resummation of the closed string
loops in the string-string scattering studied in ACV leads to an
effective geometry that depends non-trivially upon the energies of
the two colliding quanta, the metric that emerges from the
resummation of open string loops considered in this paper (e.g the
length $R_p$ defined in~\eqref{eqparam}) depends only on the D-brane
system.

Finally, we note that in our approach the existence of a non-trivial
background is taken into account by the inclusion of surfaces with
boundaries in the perturbative series, rather than by a direct
modification of the couplings in the closed string sigma model.
Since the string-brane scattering amplitude is evaluated using this
microscopic definition of the system, in principle we can also
analyse the background generated by $N$ Dp-branes in regions where
the effective description~(\ref{metri}) and~(\ref{dil}) is not
reliable.

The rest of the paper is organised as follows. In Section
\ref{openstringloops} we discuss the high-energy Regge limit of the
two-point amplitude of a massless closed string state in the
background of $N$ Dp-branes. We analyse in detail the disk and the
annulus diagrams both in momentum and in impact-parameter space and
then, relying on the results of~\cite{Amati:1987wq}, we show that
the resummed scattering amplitude has an operator eikonal form. We
also compute explicitly the first subleading correction to the
eikonal.  In Section \ref{string} we analyse the S-matrix by a
semiclassical saddle point approximation at large impact parameters
where one can expect that the description of the string-brane system
in terms of a smooth effective background metric should be
recovered. In Section \ref{deflea} we show that there is indeed full
agreement with the results obtained by a semiclassical analysis of
the propagation of a closed string in the extremal p-brane
spacetime. In particular, we show that the string computation
reproduces the classical deflection angle of a null geodesic to
next-to-leading order as well as the inelastic amplitude for the
string excitations due to the gravitational tidal forces. In Section
\ref{conclusions} we discuss, with the help of a phase diagram, the
different regimes that have been (or can be) explored in our
approach and comment on the relevance of this framework to address
some aspects of black hole physics. Finally, in Appendix~\ref{appA}
and~\ref{appC} we provide some more details about the string annulus
calculation and the corresponding field theory diagrams discussed in
Section \ref{openstringloops}. In Appendix~\ref{appB} we do the same
for the effective action (discussed in Section \ref{deflea}) that
describes the excitation of a closed string moving in the curved
spacetime sourced by the Dp-brane system.

\sect{Scattering of a closed string on $N$ Dp-branes}
\label{openstringloops}

Closed string two-point amplitudes in the background of $N$
Dp-branes can be evaluated in perturbation theory summing over
Riemann surfaces with an arbitrary number of boundaries and handles.
In order to isolate the backreaction of the branes on the geometry
of the spacetime one can consider the limit in which the open
string coupling $\lambda = g N$ is kept fixed while $g \rightarrow
0$ and $N \rightarrow \infty$. The leading contribution then comes
from surfaces with an arbitrary number of boundaries and no handles
and the resummed amplitude is expected to allow an interpretation in terms of  the propagation
of a closed string in the curved spacetime of an extremal
Dp-brane system, see Eq.~\eqref{metri}.

Our setup is reminiscent of the configuration considered
in~\cite{Fabbrichesi:1991aa}, where a highly energetic massless
closed string was used to probe a collection of massive closed
strings. The large number of constituents in the target plays in
this case the same role as the number of D-branes in our case and
can be used to suppress the contributions from the closed string
loops with respect to the tree diagrams. Of course there are
important conceptual and technical differences. For instance, a
system of $N$ parallel D-branes preserves 16 supercharges; thus it
cannot decay and the various constituents do not interact among
themselves. This allows for a precise comparison between the results
obtained from the string scattering amplitudes and those derived
from the effective p-brane metric.

In this paper we focus on the scattering of a massless closed
string belonging to the NS-NS sector. The incoming and the outgoing
strings are characterized, respectively, by their momenta $p_1$ and
$p_2$ and their polarization tensors $\e_1$ and $\e_2$. The relevant
vertex operators in the $(-1,-1)$ and $(0,0)$ pictures are
\be V_{-1}(p,\e) =
\frac{\kappa}{2 \pi}
\e_{\m \n} \, e^{-\phi - \bar{\phi}} \,
\psi^\m \, \bar{\psi}^\n \, e^{i p X} \ , \nb \ee \be V_{0}(p,\e) =
- \frac{\kappa}{\pi \ap}
\e_{\m \n} \, \left( i \p X^\m + \frac{\ap}{2} p
\psi \psi^\m \right )
 \left( i \bar \p X^\n + \frac{\ap}{2} p \bar{\psi} \bar{\psi}^\n
 \right ) e^{i p X} \ , \label{dd1}
\ee where
$\kappa^2 = 2^6 \pi^7 \ap^4 g^2$ is the gravitational coupling constant
in ten dimensions and
we follow the
conventions of~\cite{Polchinski:1998rr}. Translation invariance
along the $p+1$ directions parallel to the brane implies the
conservation of energy and momentum along the D-brane world-volume
\begin{eqnarray}
(p_{1} + p_{2})_{\parallel} = 0 \label{momcon} \ .
\end{eqnarray}
The scattering amplitude is a function of two
invariants of the residual Lorentz group $SO(1,p) \times SO(9-p)$.
Taking, without loss of generality,   $p_1$ and $p_2$ to have vanishing
components parallel to the brane, these can be written as
\begin{eqnarray}
s =  E^2 =  | k_1|^{2} =  | k_2|^{2} \ , \hspace{2cm}
t   = - (k_1 +k_2)^{2}  = - 4E^2 \sin^2 \frac{\Theta}{2} \ ,
\label{Et}
\end{eqnarray}
where $k = p_{\perp}$ denotes the $(9-p)$-dimensional vector
transverse to the world-volume of the Dp-brane and $\Theta $ the
angle between $k_1$ and $- k_2$. The two invariants correspond,
respectively, to the energy of the incoming string and to the
momentum transferred to the brane. For small deflection angles,
$k_1$ and $k_2$ are nearly (anti)parallel and their difference,
$(k_{1} - k_{2})$, defines a privileged transverse direction, that
of the large external momenta. The impact-parameter vector $b$,
which lies in the direction of $q \equiv -(k_{1} + k_{2})$, is
approximately orthogonal to such a direction and, together with it,
defines the collision plane. In the following we shall often use
bold characters to denote $(8-p)$-dimensional vectors transverse
both to the brane worldvolume and to the direction of $(k_{1} -
k_{2})$.

Besides taking the large-$N$ limit at fixed $g N$, we will also
consider this process in the high-energy, small scattering angle limit
defined by the conditions
\begin{equation}
\label{rl}
 \frac{\ap s}{\hbar} \gg 1
\ , \hspace{3cm}  \Theta \ll 1 \ ,
\end{equation}
where we have re-introduced momentarily Planck's constant to emphasize
that this limit allows for a semiclassical treatment.  As already
stressed in~\cite{Amati:1987wq}, this regime, even if
$\Theta$ is kept fixed and therefore $t$ becomes large, is dominated
by soft dynamics, i.e. by the exchange of many gravitons whose
individual transverse momenta are of order (and somewhat smaller than)
$R_p^{-1} \ll \sqrt{-t}$ and the corresponding relevant impact
parameters $b$ are somewhat larger than $R_p$.  This is the reason why
we can justify the use of the Regge limit, order by order in the loop
expansion, for the calculation of the scattering amplitude in $b$
space. After resummation at fixed $b$, we can go back to $t$ and
verify that we can cover the kinematical region~(\ref{rl}).

Note that
by fixing $g N$ we are also fixing, through eq.~(\ref{eqparam}), the
ratio $R_p/l_s$. In the rest of this paper we will consider brane configurations
with $R_p > l_s$, the stringy regime
$R_p < l_s$ will be briefly mentioned only in Section \ref{conclusions}.
When $R_p > l_s$ the string-brane interactions
are dominated by gravity effects and one expects a more transparent relation
with the effective geometry in Eqs. $(\ref{metri})$, $(\ref{dil})$.
Since the small-angle condition requires $b \gg R_p$, we can see that our scattering amplitudes
will be evaluated at impact parameters much larger than the string scale, $b \gg l_s$.
The conditions~(\ref{rl}) also imply
\begin{equation}
\label{rl1}   \frac{R_p E}{\hbar} \gg 1 \ , \hspace{3cm}
\frac{b E}{\hbar} = \frac{J}{\hbar} \gg 1 \ ,
\end{equation}
where $J$ is the (conserved) angular momentum.

Finally, we will impose an upper limit on $E$ such that the effects of
closed string loops can be neglected.  As we shall discuss more
quantitatively at the end of this section, such a limit can be
arbitrarily large if we go to sufficiently large $N$ (and thus
sufficiently small $g$). When this condition is fulfilled, the two-point amplitude
can be written as follows
\be
A(p_1, \e_1; p_2, \e_2) = i \, (2 \pi)^{p+1} \d^{p+1}(p_{1,
\parallel} + p_{2, \parallel}) \sum_{h=1}^\infty A_h(p_1, \e_1; p_2,
 \e_2) \ ,
 \label{bseries}
 \ee
where the label $h$ counts the number of boundaries.  Furthermore,
in our  limit the dependence of the scattering
amplitude on the polarization tensors simplifies considerably, the
leading term being proportional to ${\rm Tr} ( \epsilon_1
\epsilon_{2}^{T} )$
\be
A_h(p_1, \e_1; p_2, \e_2) \sim {\rm Tr} (
\epsilon_1 \epsilon_{2}^{\rm T} ) {\cal A}_h(s, t) ~ , \ee where ${\cal
A}_h$ contains only the part of the string amplitude that diverges
at high energy. Setting \be {\cal
A}(s, t) = \sum_{h=1}^\infty {\cal A}_h(s, t) \ , \label{bseries2}
\ee
we obtain the following expression for the Regge limit of the
T-matrix, related to the S-matrix by $S=1+iT$, \be \label{tmatr}
T(p_1,\e_1;p_2,\e_2) \sim (2 \pi)^{p+1} \d^{p+1}(p_{1,\parallel} +
p_{2, \parallel}) \ {\rm Tr} ( \epsilon_1 \epsilon_{2}^{\rm T} ) \
\frac{{\cal A}(s, t)}{2E} \ , \ee where the factor $2E$ takes into
account the correct normalization of the asymptotic states. The first
term in the series in Eq.~(\ref{bseries}) is the familiar disk
amplitude, the second one is the annulus amplitude. We shall first
analyse the Regge limit of these two amplitudes and then explain how
to obtain the leading behaviour of the full perturbative series
(\ref{bseries}). As in~\cite{Amati:1987wq}, the resummed
amplitude takes a simple operator eikonal form in impact parameter
space.  The eikonal operator will allow us to study both the emergence
of classical gravity effects from open string loops and the string
corrections to the classical dynamics.

\subsection{Disk and annulus amplitudes}
\label{disk}

The amplitude describing the elastic scattering of a massless NS-NS
closed string state on the disk is well-known
\cite{Garousi:1996ad,Klebanov:1995ni,Hashimoto:1996bf}\footnote{Disk amplitudes with both open
and closed strings were first computed in Ref.~\cite{Ademollo:1974fc}.} and reads
\begin{equation}
A_1(p_1,\e_1;p_2,\e_2) = - \frac{\pi^{\frac{9-p}{2}}
R_{p}^{7-p}}{\Gamma(\frac{7-p}{2})} \, {\cal K}(p_1, \e_1; p_2,
\e_2) \, \frac{\Gamma \left(- \alpha' s \right) \Gamma \left( -
\frac{\alpha'}{4}t \right)}{\Gamma \left(1- \alpha' s
-\frac{\alpha'}{4}t \right)}~, \label{A1}
\end{equation}
where the full kinematical factor ${\cal K}(p_1, \e_1; p_2, \e_2)$ can
be found, for instance, in~\cite{Garousi:1996ad}. The amplitude
exhibits poles both in the $t$-channel and in the s-channel. The
former are due to the exchange of closed strings and appear when the
two vertex operators approach each other, the latter are due to the
exchange of open strings and appear when the two vertex operators
approach the boundary of the disk. In the Regge limit the leading term
in the kinematical factor is
\be
{\cal K}(p_1, \e_1; p_2, \e_2) \sim {\rm Tr} \left(
\epsilon_1 \epsilon_{2}^{\rm T}\right) (\alpha '  s)^2 \ .
 \ee
Even if in this paper we focus on states belonging to the NS-NS
sector, we note that a similar result also holds for the elastic amplitude with
two R-R massless fields.

From~\eqref{A1} we obtain in the Regge limit
\begin{eqnarray}
{\cal A}_1(s,t) = \frac{\pi^{\frac{9-p}{2}}
R_{p}^{7-p}}{\Gamma(\frac{7-p}{2})} \, \Gamma \left( -
\frac{\alpha'}{4}t \right) e^{-i \pi \frac{\ap t}{4}} (\ap
s)^{1+\frac{\ap t}{4}} \ . \label{T1}
\end{eqnarray}
The previous formula shows that the amplitude is dominated by the
exchange in the $t$-channel of the Regge trajectory of the graviton.
The imaginary part takes into account inelastic processes where the
closed string excites open string degrees of freedom attached to the
brane worldvolume.  As the energy increases the approximation of
single-reggeon exchange eventually breaks down and the tree-level
amplitude violates unitarity. As it is well-known both in field
theory~\cite{Abarbanel:1969ek} and in string
theory~\cite{Amati:1987wq, Muzinich:1987in}, unitarity
is recovered by taking into account multi-reggeon exchanges, described
in our case by diagrams with a higher number of boundaries.

Let us consider now the amplitude with two boundaries, the second term
in the perturbative series in Eq.~\eqref{bseries}.  In general,
perturbative string amplitudes are given in terms of correlation
functions of the worldsheet theory integrated over the moduli space of
Riemann surfaces with punctures. In the simple case of a surface with
two boundaries and two punctures we have, besides  the insertion
points of the two vertex operators, $z_1$ and $z_2$, a single purely
imaginary modular parameter $\t_{\rm op} = i \t_2$.  Depending on
whether the world-sheet time direction is chosen parallel or
orthogonal to the two boundaries, this amplitude can be interpreted
either as a one-loop diagram of open strings, the annulus diagram, or
as a tree-level diagram of closed strings, the cylinder diagram.  The
two descriptions are connected by the modular transformation $\t_{\rm
op} \mapsto \t_{\rm cl} = - \frac{1}{\t_{\rm op}}$.

As we will show in the following, at large impact parameters the
dominant contributions to the scattering amplitude come from the
region of large $\l = {\rm Im}(\t_{\rm cl})$. For this reason it is
more convenient to represent the amplitude by means of two closed
string vertices and two boundary states~\cite{island:1999} and to
display its explicit form in the closed string channel
\begin{eqnarray}
A_2 (p_1, \e_1;p_2, \e_2)  = {\cal{N}} \int d^2 z_1 \, d^2 z_2
\langle B | V_{1} (z_1 , {\bar{z}}_1)  V_{2}
(z_2 , {\bar{z}}_2) D | B \rangle
\ ,
\label{Am}
\end{eqnarray}
where ${\cal{N}}$ is a normalization factor and $D$ is the closed
string propagator.  The final result is
(see~\cite{Pasquinucci:1997di,Lee:1997gwa} for a derivation of the
amplitude and Appendix~\ref{appA} for our conventions)
\be
{\cal A}_2(s, t) = \frac{\pi^3 (\ap s)^2 }{\Gamma^2\left (
\frac{7-p}{2} \right )} \frac{R_{p}^{14-2p}}{(2
\ap)^{\frac{7-p}{2}}}
\left[2 \int_{0}^{\infty} \frac{d \l}{\lambda^{ \frac{5-p}{2}}} \,
\int_{0}^{\frac{1}{2}} d \rho_1 \int_{0}^{\frac{1}{2}} d \rho_2
\int_{0}^{1} d \omega_1 \int_{0}^{1} d \omega_2 \
{\cal I}\right]  \ . \label{aa2}
\ee
In the previous formula the dependence on the external momenta is
contained in the function \be {\cal I} = e^{- \ap s V_s - \frac{\ap
t}{4} V_t } \ , \ee which is the correlation function on the annulus
of the exponential part of the vertex operators in Eq.
(\ref{dd1}). The functions $V_s$ and $V_t$ can be expressed in
terms of the Jacobi theta function $\theta_1(z|\tau)$ (see Eq.~\eqref{thet1}
for the definition)
and read
\begin{eqnarray}
V_s  &=& -  {2 \pi} \lambda  \rho^{2} + \log \frac{\theta_1 ( i
\lambda (\zeta+ \rho) | i \lambda) \theta_1 ( i \lambda( \zeta -
\rho)| i \lambda)  }{ \theta_1 (i\lambda \zeta + \omega | i \lambda)
\theta_1 (i \lambda \zeta - \omega| i \lambda) } \ , \label{Vsabla}
\\ V_t  &=& 8 \pi \lambda \rho_1 \rho_2 + \log \frac{\theta_1 (i
\lambda \rho + \omega | i \lambda) \theta_1 (i\lambda \rho - \omega|
i \lambda)  }{ \theta_1 (i \lambda \zeta +  \omega | i \lambda)
\theta_1 (i \lambda  \zeta - \omega| i \lambda) } \label{Vtabla} \ ,
\end{eqnarray}
where we introduced the variables $\rho = \rho_1 - \rho_2$, $\zeta =
\rho_1 + \rho_2$ and $\omega = \omega_1 - \omega_2$. The amplitude
does not depend on $\s = \w_1 + \w_2$ as a consequence of its
invariance under translations in the direction parallel to the
boundaries of the cylinder.

\subsection{The Field Theory limit at high energies}
\label{reggeFT}

Before presenting the detailed derivation of the high-energy limit
of the string amplitudes, which involves several technical points,
it is useful to consider the disk and annulus diagrams in the field
theory limit. This limit, which corresponds to sending $\alpha'
\rightarrow 0$ while keeping $R_p$ fixed, considerably facilitates
the analysis and allows us to illustrate our main results in a
simpler setting. The amplitudes discussed below can also be derived
by evaluating directly the Feynman diagrams contributing to the
scattering process, as described in Appendix~\ref{appC}.

The field theory limit of the disk amplitude is easily obtained from
Eq.~(\ref{T1}) and reads
\begin{eqnarray}
{\cal{A}}_1  (s, t) \rightarrow \frac{4 \,
\pi^{\frac{9-p}{2}} R_{p}^{7-p}}{\Gamma ( \frac{7-p}{2})} \ \frac{s}{(-t)} \ .
\label{T1ft}
\end{eqnarray}
As expected, at high-energy the amplitude is dominated
by the exchange of a single graviton in
the $t$-channel.

The field theory limit of the annulus diagram in~\eqref{aa2} is more
subtle since this amplitude is given in terms of an integral over the
variables $\lambda$, $\rho$, $\zeta$ and $\omega$.  It turns out that
the variable $T = \ap \lambda$ plays the role of the Schwinger
parameter of the field theory diagrams and thus has to be kept fixed
in the limit. Since $\ap \rightarrow 0$, this implies that the
relevant region of integration is the region of large values of
$\lambda$. We can then rewrite Eq.~\eqref{aa2} in the following way
\begin{eqnarray}
{\cal{A}}_{2} (s,t) \rightarrow  \frac{\pi^3 s^2
}{ \Gamma^2\left ( \frac{7-p}{2} \right )}
\frac{R_{p}^{14-2p}}{2^{\frac{7-p}{2}}} \, \left[
 \int_{0}^{\infty} \frac{d T}{ T^{ \frac{5-p}{2}}} \,\,
 \int_{0}^{1} d \zeta  \int_{-\zeta}^{\zeta} d \rho
 \int_{0}^{1}  d \omega \,\, e^{-
\ap s V_s - \frac{\ap t}{4} V_t }    \right] \ ,
\label{A2cc}
\end{eqnarray}
where now $V_s$ and $V_t$ stand for the asymptotic
form at large values of $\lambda$ of the functions in Eqs. \eqref{Vsabla} and
\eqref{Vtabla}, given in Eq.~(\ref{VsVtappr}).
The first term in $V_s$
in~(\ref{VsVtappr}) shows that at high energy the integral over $\rho$
is dominated by the region of integration around $\rho \sim0$ and
therefore we can make the further approximation
\begin{eqnarray}
\ap V_s \sim - 2 \pi T \rho^{2}~~, \hspace{2cm} \ap V_t \sim  - 2 \pi
T \zeta (1 - \zeta) \ . \label{VsVtftlim}
\end{eqnarray}
Substituting in Eq. \eqref{A2cc}, we see that the dependance on
$\ap$ disappears from the annulus amplitude. We note that in this
limit Eq.~\eqref{A2cc} agrees precisely with the field theory
integrand for the first Feynman diagram in Figure~\ref{feyn1} of
Appendix~\ref{appC}, evaluated around the saddle point $y \sim 0$ in
the Schwinger parametrization (see Eqs.~\eqref{d3a}
and~\eqref{d3ai}).

In order to make the integral over $\rho$ convergent we can continue
it analytically to negative values of $E^2$ or equivalently we can
rotate the integration contour in $\rho$ from the positive real axis
to the positive imaginary axis
\begin{eqnarray}
\int_{-\zeta}^{\zeta} d \rho \,\, {\rm e}^{2 \pi s T \rho^2}
\sim
\frac{i}{ \sqrt{2T} E} \ . \label{inte41}
\end{eqnarray}
Evaluating the two remaining integrals over $T$ and $\zeta$ we obtain
\begin{eqnarray}
{\cal {A}}_{2}^{(3)} (s, t) \rightarrow \frac{    \pi^{
\frac{9-p}{2}} E^4 }{ \Gamma^2\left ( \frac{7-p}{2} \right )}
\frac{R_{p}^{14-2p}}{2^{\frac{7-p}{2}}} \ \frac{i \sqrt{\pi}}{
\sqrt{2} E}  \ \left( \frac{2}{|t|}\right)^{\frac{p-4}{2}} \
 \Gamma (\frac{p-4}{2})
\ \frac{\Gamma( \frac{6-p}{2}) \Gamma( \frac{6-p}{2}) }{\Gamma
(6-p)} \label{ftlim3} \ ,
\end{eqnarray}
where the upper index $(3)$ indicates that this term diverges as $E^3$. An
alternative expression for the previous amplitude, which will be
useful in comparing it with the corresponding full
string theory result, is the following
\begin{eqnarray}
{\cal A}_{2}^{(3)} (s, t) =  \frac{i}{4E} \int \frac{d^{8-p} {\bf k}}{(2
\pi)^{8-p}} \ \left( \frac{ 4  E^2 R_{p}^{7-p}
\pi^{\frac{9-p}{2}}   }{ \Gamma (\frac{7-p}{2} ) }   \right)^2 \
 \frac{1 }{ {\bf k}^2 ( {\bf q}  - {\bf k})^2}~~, \hspace{1cm} t \equiv -{\bf q}^2 \ .
\label{cut6}
\end{eqnarray}
We see that ${\cal{A}}_{2}^{(3)} $ can be written as the convolution in momentum space
of two tree-level amplitudes ${\cal{A}}_1$.

The individual perturbative amplitudes, like ${\cal{A}}_1$ and ${\cal{A}}_{2}^{(3)}$,
are  divergent for large values of $E$ and therefore
at sufficiently high energy they will violate unitarity. It turns
out, however, that the leading contributions in energy coming from the disk, the
annulus and the higher loop diagrams, corresponding to surfaces with
more than two boundaries, are the terms of an exponential series and combine together
to give just
a phase in the S-matrix. This is most directly seen if we express the
scattering amplitude as a function of the impact parameter $b$ instead
of $t$. The relation between the S-matrix and the amplitudes
computed above is\footnote{We omit here for simplicity of
notation the terms
containing the polarizations
of the two external particles and the  $\delta$-function required by momentum
conservation. The complete expression is given in Eq.
(\ref{tmatr}).}
\begin{eqnarray}
S = 1 + iT = 1 + i  \frac{{\cal{A}}}{2E} \label{Amatrix} \ ,
\end{eqnarray}
while momentum and impact-parameter space are related by a Fourier transform
\begin{eqnarray}
T  ( E, b) = \int \frac{d^{8-p} {\bf q} }{ (2 \pi)^{8-p}} \, {\rm e}^{i {\bf b}
\cdot {\bf q} } \, T ( E, t) \ . \label{foutra}
\end{eqnarray}
Using the following equation
\begin{eqnarray}
\int \frac{d^D k}{(2\pi)^D} {\rm e}^{i k\cdot b} (k^2)^{\nu} = -
\frac{\nu}{\nu + \frac{D}{2}} \ \frac{2^{2 \nu}}{\pi^{D/2}}
\ \frac{\Gamma ( \nu + \frac{D}{2} +1)}{\Gamma (1-\nu)}
\frac{1}{(b^2)^{\nu + \frac{D}{2}}} \ , \label{forb}
\end{eqnarray}
we can transform both Eqs.~(\ref{T1ft}) and~(\ref{ftlim3}) to
impact-parameter space, getting\footnote{Here and many times in the
following we encounter expressions which exhibit singularities at
particular values of $p$. The correct result is obtained by taking the
limit from non-integer $p$, as in dimensional regularization, and in
throwing away irrelevant infinities.  As an example, at $p=6$
eq.~(\ref{T1T2}) produces logarithms of $b$ besides an unobservable
infinite Coulomb phase.}
\begin{eqnarray}
i T_1 (E,b) =  i \left(\frac{R_{p}^{7-p} \sqrt{\pi }E } { 2
\,b^{6-p}} \frac{ \Gamma (\frac{6-p}{2})}{
 \Gamma (\frac{7-p}{2})}   \right) \ , \hspace{0,4cm} iT_{2} (E, b)   = - \frac{1}{2}
\left(\frac{R_{p}^{7-p} \sqrt{\pi }E } { 2 \,b^{6-p}} \frac{ \Gamma
(\frac{6-p}{2})}{
 \Gamma (\frac{7-p}{2})}   \right)^2 \ .
\label{T1T2}
\end{eqnarray}
In conclusion, we obtain
\begin{eqnarray}
S (E, b) = 1 + i T_1 (E, b) + i T_2 (E, b) + \dots   = 1 + i T_1 (E, b) -
\frac{1}{2} \left( T_1 (E, b) \right)^2 + \dots  \, ,
\label{sum12}
\end{eqnarray}
which is consistent with the fact that the unitarity violating disk amplitude
exponentiates giving just a phase  in the S-matrix
\begin{eqnarray}
S (E, b) = {\rm e}^{i T_1 (E,b)} \, .
\label{sum123}
\end{eqnarray}
We can now proceed to compute the next-to-leading term in energy in
the field theory limit. In this case, as discussed in
Appendix~\ref{appC}, we need to take into account several Feynman
diagrams. Their combination reproduces the $\ap \to 0$ limit of Eq.
\eqref{eq93} in Appendix~\ref{appA} \be {\cal A}_2(s, t) \rightarrow
\frac{\pi^3 s}{\Gamma^2\left ( \frac{7-p}{2} \right )}
\frac{R_{p}^{14-2p}}{2^{\frac{7-p}{2}}} \frac{t}{4 }
\int\limits_0^\infty \frac{d T}{T^{\frac{5-p}{2}}}
\int\limits_{0}^{1} d \z \left[e^{-2\pi T
\z(1-\z)}\right]^{-\frac{t}{4} } ~. \label{ftl14} \ee
This result should be compared with the field theory integrand for
diagram (a) in Figure~\ref{feyn1} of Appendix~\ref{appC}, where we
keep the subleading term in the expansion around the saddle point in
the Schwinger parameter $y$. After performing the integrals over $T$
and $\zeta$ in Eq. \eqref{ftl14} we obtain
\begin{eqnarray}
{\cal{A}}_{2}^{(2)} (s,t) = - \frac{\pi^{ \frac{9-p}{2}}
E^2}{\Gamma^2\left ( \frac{7-p}{2} \right )}
\frac{R_{p}^{14-2p}}{2^{7-p}|t|^{\frac{p-5}{2}}} \  \Gamma
\left(\frac{p-3}{2}\right)\; B\left(\frac{5-p}{2},\frac{5-p}{2}\right)
\ . \label{A22}
\end{eqnarray}
Using again Eq.~(\ref{forb}), the previous expression
becomes in impact parameter space
\begin{eqnarray}
i T_{2}^{(1)} (E, b) \equiv i\frac{{\cal A}^{(2)}_2}{2 E}
= i\frac{\sqrt{\pi} E R_{p}^{14 -2p}}{
b^{13-2p}}  \frac{\Gamma ( \frac{13-
2p}{2}) }{4 \Gamma (6-p)} \ . \label{T2next}
\end{eqnarray}
The upper indices $(2)$ and $(1)$ denote that
${\cal{A}}_{2}^{(2)}$ and $T_{2}^{(1)} $ diverge respectively as
$E^2$ and $E$ at high energy.
In the next subsection we will
generalize these results in order to include the string corrections.

\subsection{String corrections to the leading eikonal}
\label{eikonal}

We now extend the results obtained in the previous subsection in two
directions.  On the one hand, we evaluate the high-energy limit of the
full string amplitude~(\ref{aa2}) in order to derive the string
corrections to the leading field theory result; on the other hand, we
determine the form of the terms in ${\cal A}_2$ that scale as $E^2$,
which are small in comparison to the leading behaviour $E^3$, but
still yield a divergent contributions to the $T$-matrix~\eqref{tmatr}.
In the previous Section we have already extracted those terms in the
field theory limit, but their derivation presented in the Appendix
includes also some (but possibly not all) string corrections.

The high-energy limit of a string amplitude can be determined by
evaluating the asymptotic behaviour of the multidimensional integral
over the world-sheet moduli using saddle-point
methods~\cite{Alessandrini:1972jy}. The leading terms are due to
critical points in the interior or on the boundary of the integration
domain, each critical point corresponding to a specific degeneration
limit of a Riemann surface with punctures.  We leave the technical
steps necessary to evaluate the saddle-points of the annulus with two
punctures in Eq.~(\ref{aa2}) to the Appendix~\ref{appA}. Here we just
give a sketch of the derivation and summarize the main results.

In the limit~\eqref{rl} there are no critical points in the interior
of the domain. The dominant contribution comes from the region of small
values of $\rho$, since we are taking the high-energy limit for the
external states, and from the region of large values of $\lambda$, since we are considering
impact parameters larger than the string scale. We shall
then expand the integrand for large $\l$ and small $\rho$ and
perform exactly the integrals over the angular variables $\omega_1$
and $\omega_2$, which implement the level matching condition.
Since we wish to focus
on the part of the S-matrix~\eqref{tmatr} that is perturbatively
divergent at high energies, we will consider only terms that are at
least of order $E^2$.
It turns
out that the expansion of the integrand around $\rho = 0$ yields a
leading contribution (indicated by ${\cal A}_2^{(3)}$) that scales as $E^3$
and a subleading contribution that scales as $E^2$
(indicated by ${\cal A}_2^{(2)}$).

The leading term ${\cal A}_2^{(3)}$ in the annulus amplitude is
precisely the one required for the exponentiation of the tree-level
amplitude into an operator eikonal form, which generalizes the field
theory result given in Eq. (\ref{cut6})  and reconciles the perturbative string expansion with
unitarity. More precisely, to all orders in $\ap$ the leading annulus
contribution has the following form\footnote{Strictly
speaking~\eqref{2rg4} is not well defined because there are
non-physical divergences, due to the poles of the gamma function in $V_2$, whenever
${\bf k}$ is such that $\ap (t_1+t_2-t)=-2,-4,\ldots$. These
divergences, however, arise from a region where the integrand is
suppressed by factors of $E$. As shown in the Appendix~\ref{appA},
the full amplitude is free of unphysical poles.}  (see Appendix~\ref{appA} for details)
\be
\frac{{\cal A}_2^{(3)}}{2 E} =
\frac{i}{2} \int \frac{d^{8-p} {\bf k}}{(2 \pi)^{8-p}} \ \frac{{\cal
A}_{1}(s,t_1)}{2 E} \ \frac{{\cal A}_{1}(s,t_2)}{2 E}  \
V_2(t_1,t_2,t) \ , \label{2rg4}
\ee
where
\be
V_2(t_1,t_2,t) = \frac{\Gamma\left [ 1 + \frac{\ap
}{2}(t_1+t_2-t) \right ]} {\Gamma^2\left [1 +
\frac{\ap}{4}(t_1+t_2-t) \right ]} \ , \label{2rv}
\ee
and
\be
{\bf q}^2 = - t \ ,  \hspace{1cm} t_1 = - {\bf k}^2 \, ,
\hspace{1cm} t_2 = - \left ( {\bf q} - {\bf k} \right )^2  \ .
\label{t1t2}
\ee
The function $V_2(t_1, t_2, t)$ is the vertex for the emission of
two reggeized gravitons derived in~\cite{Amati:1987wq} in the
context of high-energy string-string collisions.  The fact that
precisely the same vertex appears in the string-brane scattering
process studied in this paper is not surprising. As in the field
theory description of the deflection of a particle by an external
potential, one expects that the string will interact with the brane
through multiple exchanges of states lying on the Regge trajectory
of the graviton. The leading term of the annulus amplitude confirms
this expectation showing that the amplitude factorizes into the
product of the vertex for the emission of two reggeized gravitons
and two disk amplitudes which encode the boundary conditions
pertaining to the brane source.

In the $\ap t \to 0$ limit the string amplitude~\eqref{2rg4} reduces
to the leading field theory result~\eqref{cut6} of
Section $2.2$ and a comparison between the two expressions helps to clarify
the effect of the string corrections. We see that the graviton pole is
replaced by the complete Regge trajectory of the graviton and that the
convolution of the tree-level amplitudes now involves a non-trivial
kernel $V_2(t_1, t_2, t)$. As it stands, it is not evident
that~\eqref{2rg4} is the second term in the expansion of a simple
eikonal form. In order to better understand the structure of the
series let us consider the leading terms provided by the higher-order
amplitudes.

For a generic amplitude ${\cal A}_h$ with $h$ boundaries, the term
with the highest power of $E$ should correspond to the exchange of $h$
reggeons and scale as $E^{h+1}$. We can then write
\be
\label{ah38}
\frac{ {\cal A}^{(h+1)}_h(s, t)}{2E} \sim \frac{1}{h!}
\frac{i^{h-1}}{(2E)^{h }} \prod_{i=1}^{h-1} \int \frac{d^{8-p} {\bf
k}_i}{(2 \pi)^{8-p}} \ {\cal A}_1(s, t_1) ...{\cal A}_1(s, t_h)
V_h({\bf k}_1, {\bf k}_2,..., {\bf k}_h) \ ,
\ee
where $t_i \equiv - {\bf k}_i^2$ for $i =1 \dots h$ and
$\sum_{i=1}^{h} {\bf k_i}= {\bf q}$ generalize the conventions
in~\eqref{t1t2}, while the vertex for the emission of $h$ reggeized
gravitons  $V_h$ ~\cite{Amati:1987wq} generalizes the $h=2$
expression in~\eqref{2rv}. To resum all the leading contributions in
Eq.~\eqref{ah38}, it is important to realise that the vertices $V_h$
have a simple representation in terms of vacuum expectation values
of string vertex operators~\cite{Amati:1987wq} \be V_h({\bf
k}_1,\ldots,{\bf k}_h) = \langle 0 | \prod_{i=1}^h
\int\limits_0^{2\pi} \frac{d \s_i}{2 \pi}
 : e^{i {\bf k}_i \hat{X}(\s_i)} :  | 0 \rangle  \ ,
\ee
where the string fields $\hat{X}(\s)$ are defined in~\eqref{Xex}.
By using this operator form for the vertex $V_h$, we can rewrite the integrand
in~\eqref{ah38} as a convolution in momentum space and then diagonalize the
full string expression going to impact parameter space
\begin{eqnarray}
&&
i \frac{{\cal A}^{(h+1)}_h(s, {\bf b}) }{2E} =
\int  \frac{d^{8-p} {\bf q}}{(2 \pi)^{8-p}} \ e^{i {\bf b} {\bf q}}
\  i  \frac{{\cal A}^{(h+1)}_h(s, t)  }{2E} \nonumber \\
&& = \frac{i^h}{h!}   \langle  0 | \prod_{i=1}^{h}
\int\limits_{0}^{2 \pi} \frac{d \sigma_i }{2 \pi}   \int
\frac{d^{8-p} {\bf k}_i}{(2\pi)^{8-p}} \frac{{\cal{A}}_1 (s, - {\bf
k}_{i}^{2})}{2E} : {\rm e}^{ i {\bf k}_i ({\bf b} + \hat{X}
(\sigma_i ) ) }:  |0 \rangle \ . \label{ah40}
\end{eqnarray}
Summing the contribution of all the $h$-reggeon diagrams we finally
obtain the leading high-energy behaviour of the series in
Eq.~(\ref{bseries2})
\be
\sum_{h=1}^\infty \frac{{\cal A}^{(h+1)}_h(s, {\bf b})}{2E} \sim
\langle 0 |\frac{1}{i} \left [ e^{2 i \hat{\d}(s, b)}
- 1 \right ] | 0 \rangle \ , \label{eo}
\ee
where
\be
2 \hat{\d}(s, b) =
\int\limits_0^{2\pi} \frac{d \s}{2 \pi}
\int \frac{d^{8-p} {\bf k}}{(2
\pi)^{8-p}} \, \frac{{\cal A}_{1} (s,-{\bf k}^2)}{2 E}
: e^{i {\bf k} ( {\bf b} + \hat{{\bf X}}(\s))} : \
= \int\limits_0^{2\pi} \frac{d \s}{2 \pi}
\frac{: {\cal A}_{1}
\left (s, {\bf b} + \hat{{\bf X}}(\s) \right ):
}{2E}    \ . \label{ep}
\ee
As anticipated, the eikonal form of the amplitude is the one expected
in a high-energy process dominated by the exchange of reggeized
gravitons between the string and the branes. The string eikonal
differs from the field theory eikonal in that it involves the full
string amplitude~\eqref{T1} and the impact parameter is shifted by the
string position operator.  It is this simple shift that takes into
account effects due to the finite string size which can become
relevant already at large impact parameters, as we shall explain in
the next Section.

We now turn to the subleading term ${\cal A}_2^{(2)}$ in the annulus
amplitude~(\ref{aa2}). This subleading contribution has the same
energy dependence as the tree-level amplitude~(\ref{T1}) and its natural
interpretation is that it induces a
renormalization of the tree-level result
\be
\label{fe} {\cal A}_1(s, t) \mapsto {\cal A}^{(2)}(s, t) \equiv
{\cal A}_1(s, t) + {\cal A}_2^{(2)}(s, t) + \ldots \ ,
\ee
where the index $(2)$ denotes that all the terms in the previous
series diverge as $E^2$ at high energy. The correction to the Regge
pole term remains non-trivial also in the field theory limit and it is
crucial to reproduce the precise form of the Dp-brane solution in
Eq.~\eqref{metri}.  For instance, in
Section~\ref{deflea} we will compute the deflection angle of a null
geodesic in the Dp-brane space-time and compare it with the string
result in the limit of large impact parameters $b$. In this example one
can see explicitly that the one-loop shift ${\cal A}_2^{(2)}$ yields a
non-vanishing contribution to the deflection angle which accounts for
the classical sub-leading corrections in $R_p/b$.

The sub-leading
contributions are computed in Appendix~\ref{appA}. In particular in
Eq.~(\ref{I2st}) we give the complete expression for the term which is dominant
in the $\ap t \rightarrow 0$ limit. It is this term that,
when multiplied with the prefactor in Eq.~(\ref{aa2}), reduces
in the field theory limit to
\begin{equation}\label{A2ft}
{\cal A}_2^{(2)} (s,t) \to - s \frac{\pi^{\frac{9-p}{2}}}{2^{7-p}}
\frac{B(\frac{5-p}{2},\frac{5-p}{2}) \Gamma(\frac{p-3}{2})}
{\Gamma^2(\frac{7-p}{2})} \frac{R^{14-2p}}{(-t)^{\frac{5-p}{2}}}~,
\end{equation}
the expression already given in Eq.~(\ref{A22}).

Let us briefly comment on the subleading terms in the higher-order
(many boundary) amplitudes.  Besides the leading terms ${\cal
A}^{(h+1)}_h$ which give the eikonal operator in Eq.~\eqref{eo},
string diagrams with $h$ boundaries should also contain contributions
of type ${\cal A}_h^{(2)}$, that is with the same energy dependence as
the tree-level amplitude~(\ref{T1}). These terms are similar to the
term ${\cal A}_2^{(2)}$ just discussed and should provide additional
renormalizations of the Regge pole term to be included
in~\eqref{fe}. Clearly there should also be several other terms
scaling with a power of the energy intermediate between ${\cal
A}_h^{(h+1)}$ and ${\cal A}_h^{(2)}$. All these terms should combine
to define a high-energy S-matrix which generalizes the leading eikonal
operator~\eqref{eo}, including both subleading classical corrections
in powers of $R_p/b$ and string corrections in powers of $\ap/b^2$.
Exponentiation of the former is necessary in order to reproduce the
classical curved spacetime expectations discussed in
Section \ref{deflea}.  In order to determine the form of the latter one
should perform a detailed analysis of the higher-order amplitudes
which is beyond the scope of this investigation. Nonetheless, some
indications about these string corrections to higher-order terms in
the $R_p/b$ expansion should be provided by generalizing our study of
string propagation in the D-brane metric~(\ref{metri})
presented in Section \ref{deflea}.

To conclude this Section let us go back to the conditions under which
we can neglect higher-genus topologies, i.e. closed string loops
generating extra handles.  Since closed strings live in the bulk any
extra closed-string loop should involve the ten-dimensional Newton
constant $G_{10} \sim \kappa_{10}^2 \sim g^2 (\ap)^4$ with no
associated power of $N$. This is the reason we have already advocated
to argue that, at sufficiently large $N$, closed-string loops can be
neglected.  In order to make the argument more quantitative let us
notice that, in analogy with the ACV case, each extra handle can
contribute to the eikonal phase at most one extra power of the
asymptotic-energy variable, which is $E$ in the present case. Hence, by
dimensional arguments, each extra handle will be associated with a
factor $\frac{G_{10} E}{b^7}$.
In general, from a surface with $n$ boundaries and $m$ handles we expect a contribution
to the eikonal phase of the form
\be
\label{extrabh}
\delta(n,m) \sim \frac{E b}{\hbar} \left(\frac{R_{p} } { b}\right)^{n(7-p)}
\left(  \frac{G_{10} E}{b^7}\right)^m \sim  \frac{E b}{\hbar}
\left(\frac{E }{M_s}\right)^m  (g N)^n g^{2m} \left( \frac{l_s}{b}\right)^{(7-p)n +7m }\, .
\ee
The first correction to the disc amplitude~(\ref{T1T2}) will thus be
\be
\label{extrah}
\delta(1,1) \sim \frac{E b}{\hbar} \left(\frac{R_{p} } { b}\right)^{7-p}
\left(  \frac{G_{10} E}{b^7}\right) \sim  \frac{E b}{\hbar}  \frac{E }{M_s}
g^3 N \left( \frac{l_s}{b}\right)^{14-p}\, .
 \ee
On the basis of~(\ref{extrah}) we see that the contribution of one extra handle is negligible if
\be
\label{E<}
\frac{E}{M_s}  \ll N~ f\left(\frac{R_p}{l_s}, \frac{R_p}{b}\right)\, ,
\ee
where $f$ is a simple function of its arguments. This condition is
obviously satisfied at fixed $R_p/l_s$ (i.e. fixed $gN$) and fixed
$R_p/b$ (i.e. fixed $\Theta$) if $N$ is taken to be sufficiently
large. It is easy to see that, under this condition, we also have
$\delta(n,m) \ll1$. Note finally that, given the above-mentioned
relation between $G_{10}$ and $\ap$, the condition~(\ref{E<}) allows
for $E$ to be arbitrarily large even with respect to the
10-dimensional Planck mass.

\sect{Analysis of the amplitude}
\label{string}

The eikonal operator derived in the previous Section
\be S(s, {\bf b}) = e^{2 i
\hat{\d} (s, {\bf b})} \ ,  \hspace{1cm}
2 \hat{\d} (s, {\bf b}) = \frac{1}{2 E}
\int \frac{d \s}{2 \pi} : {\cal A}_{1}
\left (s, {\bf b} + \hat{{\bf X}}(\s) \right ): \ , \label{sb}
\ee
gives the leading behaviour, at high energy and in a series expansion
in powers of $R_p/b$, of the scattering amplitude of a massless closed
string on a stack of $N$ Dp-branes.  This amplitude, which resums the
dominant contributions of an infinite number of open string loops in
Minkowski space, should be interpreted as describing the semiclassical
propagation of a closed string in a curved spacetime, the background
generated by the D-branes.

In this Section we shall use Eq.~\eqref{sb} to analyse the way in
which this curved background influences the motion of our probe string
and to obtain some information on the background itself.  In our
approach we do not assume any knowledge of the effective metric
generated by the D-branes, rather we see it emerging dynamically from
the resummation of the open string loops.  The fact that all-order
perturbative string computations in flat space give rise to effects
which can be interpreted as due to the motion of the string in an
effective curved background was first discussed in the context of
string-string collisions at transplanckian
energies~\cite{Amati:1987wq, Gava:1989wh}.
As explained in the introduction, our results provide a simpler
instance of the same phenomenon.

Since the process we are considering is the scattering of an extended
object in an asymptotically flat spacetime, there are two main
effects we can study: the deflection of the trajectory of the
projectile and the excitation of its internal degrees of freedom. As
we will show, these two effects are neatly taken into account by the
eikonal operator~\eqref{sb}.

In the next Section we will provide a direct derivation of the
deflection angle and the excitation spectrum of a string in the
extremal p-brane background of Eqs.~$(\ref{metri})$, $(\ref{dil})$, finding
precise agreement with the results of this Section. The agreement
between the resummation of open string loops in Minkowski space and
the quantization of the string in an external metric confirms that,
for large $R_p$, a collection of $N$ coincident Dp-branes is well
approximated by the extremal p-brane solution of the supergravity
equations of motion. It also shows that, at least in certain limits,
our approach provides a quantitative tool to study the classical
dynamics of a string in a curved spacetime.  The advantage of the
present approach, based on the microscopic and manifestly unitary
D-brane description, is
that it does not rely on the existence of an effective external metric
and can be used to analyse other interesting dynamical regimes, as
discussed in Section \ref{conclusions}.

In order to derive from~\eqref{sb} both the deflection angle and the
excitation spectrum, we expand the eikonal phase in a power series in
the string position operators $\hat{X}^i$
\be
2 \ \hat{\d}(s, {\bf b} + \hat{\bf X}) \sim
\frac{1}{2 E} \left[ {\cal  A}_1(s,b)+
\frac{1}{2} \frac{\p^2 {\cal A}_1(s,b)}{\p b^i \p b^j} \
\overline{\hat{X}^i \hat{X}^j } + ... \right]
\ , \label{eos}
\ee
assuming $b \gg R_p \gg l_s \sqrt{\ln(\ap s)}$ and keeping only the
first two non-trivial terms.  The indices $i$, $j$ label the $8-p$
directions of the impact parameter space transverse to the brane and
to the collision axis and the symbol $\overline A $ denotes the
average of a local operator $A(\s, \t)$ on the worldsheet
\be
\bar{A} \equiv
\frac{1}{2 \pi} \int_0^{2 \pi} d \s : A(\s, \t = 0) : \ .
\ee
The two terms in~\eqref{eos} will give rise respectively to the
leading contribution to the deflection angle and to the leading
contribution to the tidal excitation of the string modes.  Let us
analyse these two effects in turn.

\subsection{Deflection angle up to next-to-leading order}

The first term in~\eqref{eos} is simply the Fourier transform of the
disk amplitude~\eqref{T1}
\be  {\cal A}_1(s,b)\sim s \, \sqrt{\pi} \,
\frac{\Gamma \left(\frac{6-p}{2} \right)} {\Gamma
\left(\frac{7-p}{2} \right)} \frac{R_p^{7-p}}{b^{6-p}} +
\frac{i \pi \sqrt{s}  }{\Gamma \left(\frac{7-p}{2} \right)} \sqrt{\frac{\pi \ap
s}{\ln \ap s}} \left ( \frac{R_p}{l_s(s)} \right )^{7-p}
e^{-\frac{b^2}{l^2_s(s)}} \ , \label{cutrg}
\ee
where $l_s(s)$ is the effective string length, the size of a
string of energy $E = \sqrt{s}$
\be
l_s(s) = l_s \sqrt{\ln \ap s} \ .
\ee
The previous formula shows that when $b \gg R_p$ and $R_p \gg l_s(s)$
the eikonal phase is predominantly real, since the absorption effects
due to the imaginary part in~\eqref{cutrg} becomes relevant only for
$b \le l_s(s)$.  In the computation of the deflection angle
$\Theta_p$ we can then approximate ${\cal A}_1$ with its real part.

The functional relation between the deflection angle and the impact
parameter can be derived evaluating the Fourier transform of
Eq.~(\ref{sb}) back to momentum space
\be S(s, t) \sim \int  d^{8-p} {\bf b} \ e^{- i {\bf b} {\bf q}} \
e^{ \frac{i}{2 E}
{\cal A}_1 (s, b)}   \ .
\label{fourier}
\ee
The integral is dominated by the saddle point
\be {\bf q} = \frac{1}{2 E} \frac{1}{b} \frac{\partial {\cal
    A}_1}{\partial b}  \ {\bf b} \ ,
\ee
and, using the small-angle relation $\Theta_p \sim
- \frac{\bf q \cdot \bf{\widehat{b}}}{E}$, we obtain
\be
\Theta_p = - \frac{1}{2 s} \frac{\partial {\cal A}_1}{\partial b}(s,b)
\ . \label{thdelta}
\ee
In our analysis in Section \ref{openstringloops}, we showed that the
annulus amplitude contains a term, ${\cal A}_2^{(2)}$, which has the
same energy dependence as ${\cal A}_1$ but is of higher order in
$R_p/b$.  To determine how these additional term enters in the eikonal
operator in Eq.~\eqref{sb} would require a more detailed study of the
subleading contributions coming from surfaces with more than two
boundaries.  It is however plausible, at least if one neglects the
effect of string corrections, that also this term will exponentiate
and contribute to the series expansion of the eikonal phase in powers
of $R_p/b$.  If we make this assumption, we can include in the real
part of the eikonal phase in~\eqref{thdelta} the one-loop
renormalization of the Regge pole obtaining
\be
{\rm Re} [{\cal A}_1 + {\cal A}_2^{(2)}] \sim s
\, \sqrt{\pi} \,  \frac{\Gamma \left(\frac{6-p}{2} \right)}
{\Gamma \left(\frac{7-p}{2} \right)} \frac{R_p^{7-p}}{b^{6-p}} +
s \ \frac{\sqrt{\pi}}{4 \Gamma\left ( \frac{7-p}{2} \right)}
\frac{\Gamma\left ( \frac{13-2p}{2} \right)\Gamma\left (
\frac{5-p}{2} \right)}{\Gamma\left ( 5 - p \right)} \
\frac{R_p^{14-2p}}{b^{13-2p}} \ . \ee
Our
string computation then leads to the following approximation for
the deflection angle of a null geodesic in the background of $N$ Dp-branes
\be
\Theta_p = \sqrt{\pi} \left [ \frac{ \Gamma\left(
\frac{8-p}{2}\right)}{ \Gamma\left( \frac{7-p}{2}\right)} \left (
\frac{R_p}{b} \right )^{7-p} + \frac{1}{2} \frac{\Gamma\left (
\frac{15-2p}{2} \right)}{\Gamma\left ( 6 - p \right)} \left (
\frac{R_p}{b} \right )^{2(7-p)} + O\left( \left ( \frac{R_p}{b}
\right )^{3(7-p)}  \right) \right ] \ . \label{stheta}
\ee
In the next Section we
will show that this result is in perfect agreement with the
deflection angle predicted by classical gravity in the extremal
p-brane background, giving direct evidence in favour of
our assumption that also the subleading term exponentiates.

\subsection{Tidal excitation at leading order}

We now turn to the excitations of the internal degrees of freedom of
the string. They are taken into account by the second term in
Eq.~\eqref{eos}, which is of leading order both in $R_p/b$ and in
$\ap/b^2$.  The higher-derivative terms in the Taylor expansion of the
eikonal phase in Eq.~\eqref{sb} lead to higher string corrections
weighted by the same leading power of $R_p/b$ but of higher order in
$\ap/b^2$ and are therefore suppressed when $b \gg l_s$.  It is the
term with two derivatives that provides the leading contribution to
the imaginary part of the eikonal phase, although it gives a
negligible correction to its real part and therefore to the deflection
angle. As we shall show it gives rise to absorption effects which,
unlike those due to the imaginary part of the tree-level amplitude in
Eq.~\eqref{cutrg}, become important already at large distances $b \gg
l_s(s)$ \cite{Amati:1987wq}.  This imaginary part of the
eikonal phase accounts for the excitation of the string under the
influence of the long-range gravitational field of the
brane~\cite{Giddings:2006vu}.

In order to study this effect we derive the explicit form of the
eikonal operator~\eqref{sb} in terms of the standard operators
$\a^i_n$, $\bar{\a}^i_n$, $n \in \mathbb{Z}$, corresponding to the
left-moving and right-moving modes in the
expansion~\eqref{Xex} of the fields $\hat{X}^i$. Since the eikonal
phase depends only on the modulus of the impact parameter, the matrix
of its second derivatives has the following simple structure
\be
\label{secder}
\frac{1}{4 \sqrt{s}} \, \frac{\p^2 {\cal A}_1(s,b)}{\p b_i \p b_j} =
Q_\perp(s,b) \ \left [ \d_{ij} - \frac{b_i b_j}{b^2} \right ] + \
Q_{\parallel}(s,b) \ \frac{b_ib_j}{b^2} \ , \ee with $7-p$
coincident eigenvalues associated with the components of the string
operator orthogonal to the impact parameter \be Q_{\perp}(s, b) =
\frac{1}{4 \sqrt{s}} \, \frac{1}{b} \frac{d {\cal A}_1(s,b)}{db} \ ,
\ee
and one
eigenvalue associated with the component parallel to it
\be
 Q_{\parallel}(s,b) = \frac{1}{4 \sqrt{s}} \, \frac{d^2 {\cal
     A}_1(s,b)}{db^2} \ .
\ee
The average of the square of the string coordinates can be
expressed in the following convenient form~\cite{Amati:1987wq}
\be
\overline{(\hat{X}^i)^2}
= \ap \ \sum_{n=1}^\infty \ \frac{1}{n} \ \left [ 2 \, T^i_{0,n} - 1
- T^i_{+, n} - T^i_{-, n} \right ] \ , \hspace{1cm} i = p+1, ..., 8
\ ,
\ee
where the operators $ T^i_{\a, n}$ are
\be
T^i_{\pm, n}
= \frac{1}{n} \a^i_{\mp n} \bar{\a}^i_{\mp n} \ , \hspace{1cm} 2
T^i_{0, n} = 1 +  \frac{1}{n}\left (\a^i_{- n} \a^i_{n} +
\bar{\a}^i_{- n} \bar{\a}^i_{n} \right ) \ ,
\ee
and satisfy the following commutation relations
\be
[T^i_{-, n} , \ T^j_{+, m}] = 2
\ T^i_{0, n} \ \d_{ij} \ \d_{n m} \ , \hspace{1cm} [T^i_{0, n} , \
T^j_{\pm, m}] = \pm \ T^i_{\pm, n}  \ \d_{ij} \ \d_{n m} \ .
\ee
The previous relations imply that the operators $ T^i_{\a, n}$ form
a $SU(1,1)$ algebra whenever they carry the same $n$ and $i$ labels,
otherwise they commute. To quadratic order in the string oscillators
the eikonal operator reads
\be e^{2 i \hat{\d}(s, {\bf b})}
\sim  e^{\frac{i}{2 \sqrt{s}} {\cal A}_1(s,b)} \prod_{n= 1}^\infty  \prod_{k
= p+1}^8 e^{ i  \frac{\ap}{n} Q_k \left( 2 T^k_{0, n} - 1 - T^k_{+,
n} - T^k_{-, n} \right)} \ ,
\ee
where $Q_k$ stands for $Q_\perp$ or $Q_\parallel$ according to whether
the direction $k$ is perpendicular or parallel to the impact parameter.
Using the identity
\be e^{x \left(
2 T_0 - T_+ - T_- \right)} =  e^{- \frac{x}{1-x} T_+} \ e^{- 2 \ln
(1 - x) T_0}
 \ e^{- \frac{x}{1-x} T_-} \ ,
\ee we can rewrite the eikonal operator in normal ordered form
\ba
e^{2 i \hat{\d}(s, {\bf b})} &\sim&  e^{\frac{i}{2 \sqrt{s}} {\cal A}_1(s,b)}
\prod_{n= 1}^\infty  \prod_{k = p+1}^8 \frac{e^{-
\frac{i \ap Q_k}{n}}}{1 - \frac{i \ap Q_k}{n}} \\ \nb &&  \prod_{n=
1}^\infty  \prod_{k = p+1}^8 e^{- \frac{i \ap Q_k}{n - i \ap Q_k}
T^k_{+, n}} e^{ - \ln \left ( 1 - \frac{i \ap Q_k}{n}\right ) \left(
2 T_{0, n}^k - 1 \right)} e^{- \frac{i \ap Q_k}{n - i \ap Q_k}
T^k_{-, n}} \ .
\ea
It is now immediate to evaluate the elastic
scattering amplitude, which is given by the vacuum expectation value
of the eikonal operator
\be
\langle 0 | e^{2 i \hat{\d}(s, {\bf b})}
| 0 \rangle \sim e^{\frac{i}{2 \sqrt{s}}  {\cal A}_1(s,b)} e^{-
i \ap \g \left ( (7 - p) Q_\perp + Q_{\parallel} \right)} \Gamma
\left (1 - i \ap Q_{\parallel} \right ) \Gamma^{7-p} \left (1 - i
\ap Q_\perp \right ) \ ,
\ee
where we used the infinite-product representation of
the gamma function and $\g$ is the Euler-Mascheroni
constant. From the previous formula one can see that the string
corrections to the real part of the eikonal phase are negligible at
large impact parameters. Their most important effect in this limit
is to induce a non-vanishing imaginary part which, after using
the identity $\Gamma(1 + i x) \Gamma(1 - i x) = \frac{\pi x}{\sinh \pi x}$,
can be written as follows
\be
\left | \langle 0 |
e^{2 i \hat{\d}(s, {\bf b})} | 0 \rangle \right | \sim e^{- \frac{1}{2
    \sqrt{s}} {\rm Im} {\cal A}_1(s,b)} \left
[ \frac{\pi \ap Q_\perp}{\sinh \pi \ap Q_\perp}\right ]^{\frac{7-p}{2}}
\left [  \frac{\pi \ap Q_\parallel}{\sinh \pi \ap Q_\parallel} \right
]^{\frac{1}{2}}  \ .
\label{tash}
\ee
At high energy the previous formula becomes
\ba
\left | \langle 0 |
e^{2 i \hat{\d}(s, {\bf b})} | 0 \rangle \right | &\sim& e^{-
  \frac{1}{2 \sqrt{s}} {\rm Im} {\cal A}_1(s,b)} (2 \pi \ap)^{\frac{8-p}{2}}
|Q_\perp(s,b)|^{\frac{7-p}{2}} \left | Q_{\parallel}(s, b) \right
|^{\frac{1}{2}}  \nb \\
&& e^{- \frac{\pi}{2} \ap \left [ (7 - p) |Q_\perp(s,b)| +  \left |
Q_{\parallel}(s, b) \right | \right ]} \ . \label{ta}
\ea
Using the explicit form of ${\cal A}_1(s,b)$ we obtain
\be
Q_\perp(s,b) \equiv Q_1(s, b) =  -  \frac{\sqrt{\pi}}{2} \sqrt{s}
\frac{\Gamma \left(\frac{8-p}{2} \right)} {\Gamma \left(\frac{7-p}{2}
  \right)} \frac{R^{7-p}}{b^{8-p}} \ , \hspace{1cm}
 Q_\parallel(s, b) = - (7 - p) \ Q_1(s, b) \ .
\ee
The expression for the elastic amplitude simplifies
\be \
\langle 0 | e^{2 i \hat{\d}(s, {\bf b})} | 0 \rangle \sim e^{\frac{i}{2
    \sqrt{s}}  {\cal A}_1(s,b)} \Gamma \left (1 + i \ap (7 - p) Q_1
\right ) \Gamma^{7-p} \left (1 - i \ap Q_1 \right ) \ , \ee
\be
\left | \langle 0 | e^{2 i \hat{\d}(s, {\bf b})} | 0 \rangle \right
| \sim e^{- \frac{1}{2 \sqrt{s}}  {\rm Im} {\cal A}_1(s,b)} \left ( 2 \pi \ap
|Q_1(s, b)| \right )^{\frac{8-p}{2}} \sqrt{7-p} \ e^{- \pi \ap (7 - p)
| Q_1(s,b) |} \ , \label{ta1}
\ee
and we can see that the absorption of
the elastic channel due to string excitations becomes non negligible
for $b \le b_D$ where
\be
\label{bD}
b^{8-p}_D = \frac{\pi}{2} \ap \sqrt{\pi s}
(7-p) \frac{\Gamma \left(\frac{8-p}{2} \right)}{\Gamma
\left(\frac{7-p}{2} \right)} R_p^{7-p} \ .
\ee
At large distances this effect is more important than the inelastic
absorption given by~(\ref{cutrg}) and becomes relevant already at
impact parameters large compared with both the string scale and the
curvature of the brane background. In the next Section we will show
that the tidal excitations of the string given by~(\ref{ta1}) agree
with the results of a semiclassical computation in the Dp-brane
spacetime.

\sect{Comparison with curved-spacetime expectations}
\label{deflea}

In this Section we will compare the results we have obtained from
string computations in Minkowski spacetime to what we would expect
from the propagation of point or string-like objects in the
non-trivial background $(\ref{metri})$, $(\ref{dil})$ generated by the
D-branes. Note that, in the high-energy limit, our probes are only
sensitive to the metric part of the background (since gravitational
couplings are proportional to the energy), while the dilaton and the
RR-form backgrounds would provide subleading corrections not
considered in this paper.

We will focus our attention on the comparison of  two effects:
\begin{itemize}
\item Deflection angles up to next-to-leading order in the
point-particle limit.
\item Tidal excitation of stringy probes at leading order.
\end{itemize}
As we shall see, there is full agreement on both effects between the
string-loop calculations and curved-spacetime expectations.

\subsection{Classical deflection up to next-to-leading order}
\label{angle}

Let us start by a (quite standard) computation of the deflection
suffered by a massless point-like probe in the metric~(\ref{metri})
produced by our stack of Dp-branes. We shall use a slightly more
general form of the metric allowing one to change conformal frame
(e.g.  from the string frame to the Einstein frame).  Since the
coordinates involved in the geodesics are only the time $t$ and the
spatial coordinates $r$ and $\theta$ of the plane in which the motion
takes place, we can limit ourselves to the following part of the
metric
 \begin{eqnarray}
ds^2  = - \alpha(r) dt^2 + \beta(r) (d r ^2 + r^2 d\theta^2) \ .
\label{1a}
\end{eqnarray}
 Since the metric does not depend on the time
$t$ and the angle $\theta$ there are two conserved quantities, the
energy $E$ and the angular momentum $J$. From these two conservation
laws and from the invariance of the action under arbitrary
reparametrizations of the world-line coordinate $u$, one can find a
differential equation relating $\theta$ and the radial coordinate
\begin{equation}
\frac{d\theta}{dr} = -\frac{b}{r^2 \sqrt{\frac{\beta}{\alpha} -
\frac{b^2}{r^2 }}} ~~ \Leftrightarrow ~~ \frac{d\theta}{d\rho} =
\frac{\hat{b}}{\sqrt{1+ \rho^{7-p} - {\hat{b}^2}{\rho^2}}}~,
\label{eqtraj}
\end{equation}
where $b = J/E$, $\rho=R_p/r$, $\hat{b}=b/R_p$.
Notice that the previous result depends only on the ratio $\alpha/\beta$ and
it is therefore invariant under an $r$-dependent rescaling of the whole
metric. This means that both the string and the Einstein frame metric
yield the same equation~(\ref{eqtraj}) for the classical
trajectory. In the last step of~(\ref{eqtraj}) we used the actual form
of $\alpha/\beta$ given in Eq.~(\ref{metri}).

From~(\ref{eqtraj}) we see that the value of the angle $\theta$ at
 the turning point $r_*$ is
\begin{equation}
\theta(r_*) = \int_\infty^{r_*} \frac{d\theta}{dr} dr =
 \int_{0}^{\rho_*} {d\rho} \frac{\hat{b}}{\sqrt{1+
\rho^{7-p} - {\hat{b}^2}{\rho^2}}} \ ,
\end{equation}
where $\rho_*=R_p/r_*$ is the smallest root of the equation $ 1+
\rho^{7-p} - {\hat{b}^2}{\rho^2}= 0$.  Since the trajectory of a probe
particle in the metric in Eq.~(\ref{1a}) is symmetric around $r_*$,
the deflection angle $\Theta_p$ is given by
\begin{equation}
\Theta_p = 2\theta(r_*) - \pi ~\Rightarrow~~~
\Theta_p  = 2  \int_{0}^{\rho_*} {d\rho} \frac{\hat{b}}{\sqrt{1+
\rho^{7-p} - {\hat{b}^2}{\rho^2}}} - \pi \ . \label{8a}
\end{equation}
The integral can be performed explicitly in terms of elementary
functions for the cases $p=5,6$ yielding
\begin{eqnarray}
\tan \frac{\Theta_6}{2} = \frac{1}{2\hat{b}} \ , \hspace{1cm}  \Theta_5 =
\frac{\pi}{\sqrt{1 -  \left(\frac{1}{\hat{b}}\right)^2}}  - \pi \ .
\label{theta6}
\end{eqnarray}
For the case $p=3$ we get instead
\begin{eqnarray}
\Theta_3 = 2 \, \sqrt{1+k_3^2} \, K (k_3) - \pi  \ , \hspace{1cm}
K(k_3) = \int_{0}^{1} \frac{ dv} {\sqrt{(1- v^2)(1- k_3^2 v^2) }} \
, \label{th3}
\end{eqnarray}
where $K$ is the complete elliptic integral of first kind and
\begin{equation}
k_3 = -1+ \frac{\hat{b}}{2} \left(\hat{b}- \sqrt{\hat{b}^2-4}\right) \ .
\end{equation}

Similar expressions involving elliptic integrals can be given for $p =
1$ and $p = 4$ while for the cases $p = 0$ and $p = 2$ we do not have
an expression in terms of special functions. Nonetheless, the leading
and next-to-leading terms in the large impact parameter expansion for
arbitrary $p$ can be computed and read
\begin{equation}
\Theta_p = \sqrt{\pi} \left[ \frac{\Gamma (\frac{8-p}{2})}{\Gamma
(\frac{7-p}{2})} \left( \frac{R_p}{b} \right)^{7-p} + \frac{1}{2}
\frac{\Gamma (\frac{15- 2p}{2})}{\Gamma ( {6-p} )} \left(
\frac{R_p}{b} \right)^{2(7-p)} + \dots  \right] \ ,
\label{leadnext}
\end{equation}
in perfect agreement with the string calculations in Eq.~\eqref{stheta}.

We should make a remark at this point about the order at which we
expect to find agreement between the string calculation and the
external metric one.  In matching the two results we have identified
the impact parameter $b$ of the string calculation, defined by the
Fourier transform~\eqref{fourier}, with the impact parameter of the
geodesic calculation, defined as $b= J/E$. We expect such an
identification to fail at order $\Theta^3_p$, when, for instance,
$\sin \Theta_p$ starts to differ from $\Theta_p$.

\subsection{Tidal excitation of the closed string at leading order}
\label{tides}

We shall now compare the results of the previous Section concerning
the possible excitation of the probe closed strings with what one
obtains by quantizing a closed string in the non-trivial metric
(\ref{metri}). A similar exercise in the case of string-string
collisions leads, to leading order, to agreement with expectations for
quantizing a closed string in an Aichelburg-Sexl metric
\cite{Veneziano:1988qq}.

In the case at hand the end result of the string calculation is the
eikonal-operator formula~\eqref{sb}.  Such a formula refers to the
leading contribution in $R_p/b$ but is supposed to hold at all orders
in $\alpha'/b^2$.  On the other hand, the curved-spacetime calculation
we shall present below is limited to small string fluctuations around
the point-particle null geodesic while, in principle, it can be
extended to higher orders in $R_p/b$. Our comparison will be made in
the overlap of the domains of validity of the two calculations, namely
at leading order both in $\alpha'/b^2$ and in $R_p/b$.  In spite of
this the perfect agreement between the two calculations appears to be
almost miraculous, given the very different techniques being used,
and represents in our opinion a highly non trivial check of the
validity of our approach.

In order to set up the curved space calculation we shall follow
\cite{Blau:2002mw} and rewrite the full metric
as
\begin{equation}
ds^2  =
\alpha(r) \left(-dt^2 + \sum_{a=1}^p (dx^a)^2\right) +
\beta(r) \left(d r ^2 + r^2 (d \theta^2 +
\sin^2 \theta  d\Omega^2_{7-p})\right) \ ,
\label{1b}
\end{equation}
where in our case $\b(r)=1/\a(r)=\sqrt{H(r)}$. This reduces
to~\eqref{1a} on the plane of the null geodesic considered in Sect.
4.1 and we are now interested in describing the metric around such a
geodesic.  This is done by first going to a system of adapted
coordinates $u, v, z, x^a, y^i$ in which the geodesic corresponds to
constant $ v, z, x^a, y^i$ and $u = u(r)$ plays the role of the
affine parameter along the geodesic
\begin{eqnarray}
dv &=& - dt + b d\theta +C dr \ , \hspace{2cm} dz =
d(\theta +\bar\theta(u))
 \ , \label{vzxy} \\
 du &=& \pm \frac{\beta dr}{C} \ , \hspace{3.6cm}
C(r) = \sqrt{\frac{\beta(r)}{\alpha(r)} - \frac{b^2}{ r^2}} \ .
\label{C}
\end{eqnarray}
Here $\bar{\theta}(u)$ is the angular coordinate $\theta$, evaluated
along the null geodesic as in~\eqref{eqtraj} and expressed in terms
of $u$ via Eq.~(\ref{C}). The physical meaning of the $7$
coordinates $x^a, y^i$ is that they represent fluctuations
orthogonal to the null geodesic and, respectively, parallel to the
directions of the brane world-volume ($x^a$) or along the $(7-p)$
directions ($y^i$) which are orthogonal both to the brane and to the
plane of the geodesic. The $z$ coordinate is orthogonal to the
brane, but lies in the plane of the geodesic. In our conventions the
point $u=0$ corresponds to the turning point $r_*$ and the choice of
sign in the equation for $u$ depends on the point of the geodesic we
are considering: we choose the minus sign in the approaching region
and so parametrize the part of the geodesic from infinity to $r_*$
with the interval $-\infty<u\leq 0$; for the remaining part from
$r_*$ to infinity we choose the plus sign and so it corresponds to
the interval $0 \leq u <\infty$. In these adapted coordinates the
metric takes the form
 \begin{eqnarray}
ds^2  = 2 du dv - \alpha dv^2 + 2b \alpha dv dz +
r^2 \a C^2 dz^2 + \alpha  dx^a dx^a
+ \beta  r^2  \sin^2(z- \bar{\theta})  d\Omega^2_{7-p}.
\label{1c}
\end{eqnarray}
At this point we can take the Penrose limit of the above metric
(corresponding to the high-energy limit for the probe) and focus on a
small neighborhood around the null geodesic by expanding to the
quadratic order the dependence on all coordinates transverse to the
light-cone and keeping only the linear terms in $v$. This clearly
eliminates the $dv^2$ and $dvdz$ terms, then by the following change
of coordinates
\begin{align}
&
z = \frac{\hat{y}^0 }{\sqrt{r^2\a C^2}} \ , \hspace{1cm}
y^i = \frac{\hat{y}^i}{\sqrt{\beta} \ r \sin \bar{\theta}} \ , \hspace{1cm}
x^a =\frac{\hat{x}^a}{\sqrt{\alpha}} \ ,
\\
\nonumber & v = \hat{v} +\frac 12 \left[ \sum_{a=1}^{p} \hat{x}_a^2
\, \partial_u\ln(\sqrt{\alpha}) + \sum_{i=1}^{7-p} \hat{y}_i^2
\, \partial_u\ln(\sqrt{\beta r \sin \bar{\theta}}) + \hat{y}_0^2 \, \partial_u\ln(\sqrt{r^2
\alpha C^2})\right] \ ,
\end{align}
we can bring the metric in the pp-wave form
\cite{Blau:2002mw}
 \begin{eqnarray}
ds^2 & =& 2 du d {\hat{v}} + \sum_{a=1}^p d\hat{x}_a^2 +
\sum_{i=1}^{7-p} d \hat{y}_i^2 + d \hat{y}_0^2 + {\cal G}(u,
\hat{x}^a, \hat{y}^i, \hat{y}^0) du^2
~~,~~ \nonumber \\
 {\cal G} &=& \frac{ \partial_u^2 \sqrt{\alpha}}{\sqrt{\alpha} }
 \sum_{a=1}^{p}  \hat{x}_a^{2} + \frac{\partial_u^2 ( \sqrt{\beta} r \sin \bar{\theta})}{
\sqrt{\beta} r \sin\bar{\theta}} \sum_{i=1}^{7-p} \hat{y}_i^2 +
\frac{\partial_u^2 \sqrt{\beta r^2 - b^2 \alpha}}{\sqrt{\beta r^2 -
    b^2 \alpha}} \hat{y}_0^2
 \
\nonumber \\ &\equiv&  {\cal G}_x \ \sum_{a=1}^p  \ \hat{x}_a^{2} +
{\cal G}_y \ \sum_{i=1}^{7-p} \ \hat{y}_i^2 + {\cal G}_0 \
\hat{y}_0^2 ~. \label{A}
\end{eqnarray}
The bosonic part of the string sigma model then reads
\begin{equation}
S = S_0 -\frac{1}{4\pi\alpha'} \int d\tau \int_0^{2\pi} \!\!\!
d\sigma~ \eta^{\a\b} ~  \partial_\a U \partial_\b U {\cal G}(U, X^a,
Y^i, Y^0)  \,,
\end{equation}
where from now on we will drop the hats on the coordinates. In the
previous equation $S_0$ is the free Minkowski string action,
$\eta_{\a\b}$ the flat worldsheet metric and ${\cal G}$ is as in
Eq.~(\ref{A}) but now considered as a function of the string
coordinates (denoted by capital letters).  In these coordinates
string quantization is quite easy if we choose (within the class of
orthonormal gauges) the light-cone gauge\footnote{The factor $2$
usually present in the r.h.s of Eq.~(\ref{Utau}) is absent here
because we take $0\leq \sigma\leq 2 \pi$. } \eq U(\sigma, \tau) =
\alpha' p^u \tau \rightarrow \alpha ' E \tau \ . \label{Utau} \eqx
This choice drastically  simplifies the non-trivial part of the
sigma model action to give \eqn && S - S_0 =  \frac{E}{2}
\int_0^{2\pi} \frac{d\sigma}{2 \pi} \int_{-\infty}^{+\infty} du ~
{\cal G}(u, X^a(\sigma, u/ \alpha' E),
Y^i(\sigma, u/ \alpha' E), Y^0(\sigma, u/ \alpha' E))  \nonumber \\
&\rightarrow&  \frac{E}{2} \int_0^{2\pi} \frac{d\sigma}{2 \pi}
\int_{-\infty}^{+\infty} du \left({\cal G}_x(u)  \sum_{a=1}^{p}
 X_a^2(\sigma, 0) +  {\cal G}_y(u)  \sum_{i=1}^{7-p} Y_i^2(\sigma, 0) + {\cal G}_0(u)
Y_0^2(\sigma, 0)\right) \nonumber \\ \label{cxy0} &\equiv&
\frac{E}{2} \int_0^{2 \pi} \frac{d\sigma}{2 \pi} \left( c_x
  \sum_{a=1}^{p}  X_a^2(\sigma, 0) +  c_y \sum_{i=1}^{7-p} Y_i^2(\sigma, 0) +  c_0 Y_0^2(\sigma, 0)
\right)  \ , \eqnx where in the second step we have used the
high-energy limit. In this way the integrals over $u$ decouple from
the string coordinates and just provide $c$-number coefficients
${\cal G}$ to the quadratic action of the fluctuations. Because of
the change of sign in $du/dr$ at $u=0$ the integrals under
consideration are twice the same integrals between 0 and $\infty$.

At first sight all the fluctuations of the closed strings, both in the
Neumann and in the Dirichlet directions, appear to be excited. It
turns out, however, that the former are {\it not} excited to the
leading order in $R_p/b$ to which we are working. This is because we
can write
\begin{equation}
 c_x = 2\int_0^\infty \frac{ \partial_u^2
   \sqrt{\alpha}}{\sqrt{\alpha} } du = 2\int_0^\infty
\left[\left({\partial_u\ln\sqrt{\alpha}}\right)^2 +
\partial_u^2\left(\ln\sqrt{\alpha} \right)\right] du~.
\label{ax}
\end{equation}
In the second integral the first term is proportional to
$\left(\frac{R_p}{b}\right)^{2(7-p)}$ and so it yields a contribution
of higher order in $R_p/b$, while the second term, being the total
derivative of a function that vanishes on the integration boundaries,
gives zero (see Appendix~\ref{appB} for more details).  Hence, to this
order, $c_x=0$.

This is not true for the other two fluctuations. The coefficient of
the $Y_i$ fluctuations is quite simple to evaluate by writing, using
again the trick that we have just used,
\begin{equation}
 c_{y} =
2\int_0^\infty
\left[\left(\partial_u\ln{\sqrt{\beta} r \sin\bar{\theta}}\right)^2 +
\partial_u^2\left(\ln{\sqrt{\beta} r \sin\bar{\theta}}
\right)\right] du ~.
\label{ay}
\end{equation}
As in the previous case, the first term does not contribute at the
order $\left(\frac{R_p}{b}\right)^{7-p}$ we are interested in, but, as
shown in Appendix~\ref{appB}, this time the $u=0$ boundary provides a
non-vanishing contribution for the second term and we obtain
$c_y = -\Theta_p/b$.

Inserting the leading value of $\Theta_p$ from~\eqref{leadnext} we see
that the coefficient of $Y^2$ agrees with the one obtained in Section \ref{string}
from the second derivative of the phase shift in the directions
orthogonal to the brane and to $\vec{b}$, i.e.
\eq
\label{cyre}
c_y = -  \frac{\sqrt{\pi}}{b} \left( \frac{R_p}{b}\right)^{7-p}
\frac{\Gamma(\frac{8-p}{2})}{\Gamma(\frac{7-p}{2})}  \Longrightarrow
\frac{E}{2} c_y = Q_{\perp} (s,b)  ~.
\eqx
Computation of the coefficient of the $Y_0$ fluctuations is a little
more involved since the analogue of the square of the first derivative
in~\eqref{ax} is not subleading: one has to perform carefully the
second derivative, expand it to leading order in $R_p/b$, and
integrate explicitly each term over $u$.  This lengthy but
straightforward exercise (reported in Appendix~\ref{appB}) gives for
the coefficient of the $Y_0$ fluctuations
\eq
c_0 =  \frac{\sqrt{\pi}}{b} (7-p)  \left( \frac{R_p}{b}\right)^{7-p}
\frac{\Gamma(\frac{8-p}{2})}{\Gamma(\frac{7-p}{2})} \Longrightarrow
\frac{E}{2} c_0 = Q_{\parallel} (s, b) \,\,,
\label{c00}
\eqx
again in perfect agreement with the string calculation.

\sect{Conclusions}
\label{conclusions}

\begin{figure}[htp]
\begin{center}
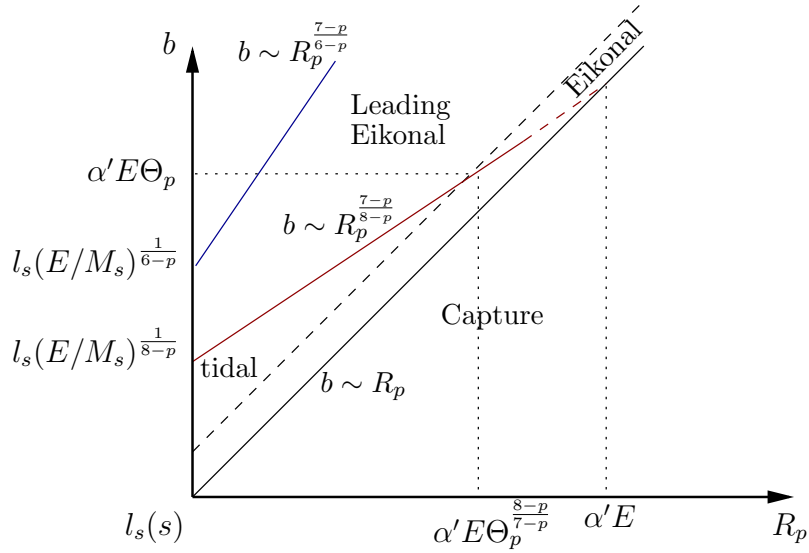
\end{center}
\caption{\label{phdiag} {\it Various qualitative
 regimes in the scattering of a closed string of fixed high energy off a
 stack of Dp-branes as a function, in a log-log plot, of $R_p$ and $b$, both
taken to be bigger than the effective string scale $l_s(s)$. The
different regions are discussed in the text.}}
\end{figure}

The eikonal operator derived in this paper allows us to describe
different regimes of the scattering of a closed string off a stack
of $N$ Dp-branes. These regimes are characterized by the relative
magnitude of the scales involved in the dynamics, namely the energy
of the colliding string $E$, the impact parameter $b$, the curvature
radius $R_p$ of the brane background and the string length $l_s$. In
analogy with ACV, it is useful to draw, as in Figure~\ref{phdiag},
the different regimes in an $(R_p,b )$ plane marking on both axis
the effective string length $l_s(s) = l_s\sqrt{{\rm ln}~\ap s}$.
Notice that, in our case, the parameter $R_p$, which sets the scale
of the effective geometry, is independent of $E$ and we can
therefore describe the various regimes at a fixed large energy $\ap
s \gg 1$. We have also limited the diagram to the region $R_p, b >
l_s(s)$ since, in this paper, we have not discussed the new
phenomena that occur for $b$ or $R_p$ smaller than $ l_s(s)$.

There are various distinct regimes that we shall now describe by
going from the top-left to the bottom-right part of
Figure~\ref{phdiag}. For very large $b$ (above the blue line) the
eikonal phase is small and we are in a perturbative -- rather than
in a classical -- regime. This region corresponds to infinitesimal
deflection angles (as $b$ becomes larger and larger). The straight
black lines at 45 degrees represent curves of constant deflection
angle: the solid one is meant to represent a critical ratio of
$R_p/b$ below which, classically, a test point particle is captured,
while the dashed one represents a typical, small deflection angle
$\Theta \ll 1$. The red line represents instead the impact parameter
$b_D$, defined in Eq. \eqref{bD}, below which tidal excitation
becomes relevant.

In the region bordered by the blue, the red and the black-dashed
line the leading eikonal approximation is reliable and elastic
unitarity holds to a very good approximation. Below the red and
above the black-dashed line we have the tidal excitations computed
in Section $2.3$, while in the complementary region (above the red
but below the black-dashed line) we need to use an improved eikonal
including higher-order classical corrections whose first term was
computed in Section $2.2$. Tidal effects, however, should be small
in this region. Clearly, as we go further down in the diagram, we
encounter a region in which both kinds of correction (string-size
and classical) come together. When the impact parameter is close to
$R_p$, we expect large corrections to the eikonal operator discussed
in this paper and, for this reason, the solid line, signaling the
onset of the string-tidal effects, ends before meeting the $b\sim
R_p$ line.

In this paper we have focused on the upper part of the diagram in
Figure~\ref{phdiag}, where the impact parameter is large compared to
$R_p$ and gravity effects dominate the interaction between the
Dp-branes and the string probe. It is in this region that the
comparison between the string dynamics in the extremal p-brane
background and the dynamics resulting from the string scattering
amplitudes is most transparent. Having tested our approach in this
regime, there are two other interesting but more difficult regions
to consider. The first is the string region $R_p < l_s(s)$, where
string corrections to the geometry are important and one expects
that the dynamics will be very different from the one predicted by
the effective background in Eqs. $(\ref{metri})$, $(\ref{dil})$. The
second is the region $b \sim R_p > l_s(s)$, where the dynamics
should be  dominated by strong gravity effects. As we lower the
impact parameter to study these two new regions, we should also be
able to make contact with the analysis of high-energy amplitudes at
fixed angle as discussed in~\cite{Gross:1987kza} and, in the context
of D-branes, in \cite{Barbon:1996ie,Bachas:1999tv}.

The quantitative analysis of a string-brane scattering process in
the string region $R_p < l_s(s)$ and in the strong gravity region $b
\sim R_p$ requires some control on both the classical and string
corrections to the leading eikonal operator. An example of the
possible effects of these corrections is provided by the string
excitations along the Dp-brane world-volume. Although these
excitations are absent in~\eqref{ep}, the analysis of
Section~\ref{deflea} suggests that they should become relevant at
smaller impact parameters. An effective way of studying these kinds
of corrections is to calculate amplitudes similar to those discussed
in this paper but involving also massive string
states~\cite{criswill}. From a more general perspective, it would be
very interesting to derive and interpret the corrections to the
eikonal operator which are of a higher order in $\frac{R_p}{b}$ and
$\frac{l_s(s)}{b}$ by computing explicitly the high-energy behaviour
of amplitudes with three or more boundaries. This kind of analysis
might be possible by using Regge-Gribov techniques
\cite{Amati:1990xe}.

Another interesting generalization of our setup is to change the
nature of the massive target and possibly also the nature of the
asymptotic space by including compact directions. For instance,
bound states of different D-branes, such as the D1/D5 system, have
been studied intensively in recent years and provide a tractable
system that has an exponential degeneracy of states (see for
instance~\cite{Mathur:2005zp} and references therein). The
geometries corresponding to these states are known and have some
interesting features such as the breaking of rotational symmetry in
the transverse space. It would be interesting to generalize the
string computations of~\cite{Giusto:2009qq,Black:2010uq} and to see
how the properties of these microstate solutions affect the
high-energy scattering considered in this paper. Furthermore, by
adding a momentum charge to the D1/D5 bound state, one obtains the
configurations studied in~\cite{Strominger:1996sh} which are related
to a black hole with a macroscopic horizon.

In all these examples the most interesting region from the point of
view of gravity and black-hole physics is that of small impact
parameters ($b\lesssim R_p$), where, as previously noted, the
eikonal phase receives large classical corrections. However, as was
the case for the simpler regime analysed in this paper, the mixed
open/closed string amplitudes which describe the collision of a
string with a D-brane configuration might be more tractable at high
energies. They could then provide an ideal framework for discussing
in a quantitative way the dynamics of matter falling beyond a
horizon and the resulting small perturbations of a black hole above
its ground state.

\vspace{2mm}
\noindent {\large \textbf{Acknowledgements} }

\vspace{2mm} We would like to thank Daniele Amati, Costas Bachas,
William Black, Marcello Ciafaloni, Emilian Dudas, Stefano Giusto,
Vishnu Jejjala, Lello Marotta, Cristina Monni, Yaron Oz and Gabriele
Travaglini for fruitful discussions.  This work is partially supported
by INFN and by STFC under the Rolling Grant ST/G000565/1.  GD would
like to thank the members of the Laboratoire de Physique Th\'eorique
de l'Ecole Normale Sup\'erieure in Paris, where part of this work was
done, for the warm hospitality extended to him.

\appendix

\sect{The saddle point in the annulus moduli space}
\label{appA}

\begin{figure}[htp]
\begin{center}
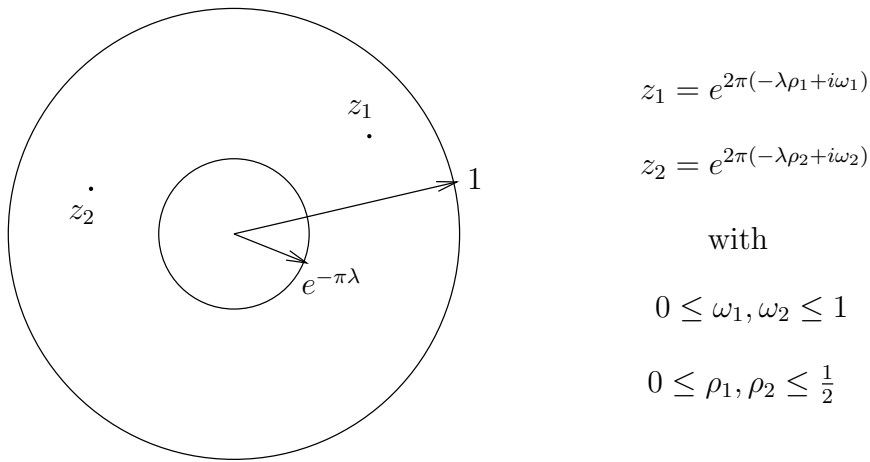
\end{center}
\caption{\label{annu} {\it The string world-sheet is represented by
the
  annulus between the outer circle of radius $1$ and the inner circle
  of radius $e^{-\pi\l}$. The two external states are represented by
  the two punctures located at $z_1$ and $z_2$.}}
\end{figure}

The string amplitude~\eqref{aa2} contains an integral over the moduli
space of a cylinder with two punctures in its interior representing
the external closed string states. Our parametrization of this surface
is summarized in figure~\ref{annu}.

We are interested in a particular kinematics where both the impact
parameter and the energy of the incident states are much larger than
the scale fixed by the string length: $\sqrt{\ap} E \gg1$ and $b \gg
l_s(s)$. In this regime the integral over the world-sheet moduli is
dominated by the contribution of a critical region at small values
of $\rho = \rho_1 - \rho_2$ and large values of $\lambda$. Indeed,
the large $b$ limit implies that the momentum exchanged between the
string state and the D-branes is small and therefore the corner of
the string integration region that is more relevant is $\lambda\to
\infty$. Moreover, we are taking the high-energy limit for the
external states and, as we will see, the integrand is exponentially
suppressed unless $\rho \sim  0$.

The Jacobi theta
function
\begin{equation}\label{thet1}
\theta_1(\nu|\tau) = -2 e^{\pi i \tau/4} \sin(\pi\nu)
\prod_{m=1}^\infty \left[ (1-e^{2\pi im \tau}) (1- 2 \cos(2\pi\nu)
e^{2\pi i m \tau} +
 e^{4\pi i m \tau}) \right] \ ,
\end{equation}
which appears in the definitions~\eqref{Vsabla} and~\eqref{Vtabla} has a simple
expansion in our critical region. We can then derive the
following expressions for the large-$\lambda$ behaviour of $V_s$ and $V_t$
\begin{eqnarray}
V_s &\sim& -2\pi \lambda \rho^2 -4 \left(\sin^2 \pi \omega + \sinh^2
\pi \lambda \rho \right ) \left( e^{- 2 \pi \lambda \zeta} + e^{- 2
\pi \lambda (1- \zeta)} \right )~ ,  \label{VsVtappr} \\
V_t &\sim&  - 2 \pi \lambda \left[\zeta (1 - \zeta) +\r^2\right] +
\ln \left[4 \sin^2 \pi \omega + 4 \sinh^2 \pi \lambda
  \rho \right]~.  \nonumber
\end{eqnarray}
In the expansion for $V_t$ we kept only the contributions that
do not have any exponential factors, while for $V_s$ we also included
the exponential terms, since they are sizable as soon as $\l > \ln \ap
s$.  The first term in each expression corresponds to the leading field
theory contribution considered in Section~\ref{reggeFT}, while the other terms yield
the sub-leading field theory contributions and the string corrections.

Let us now focus on the integral in the square parenthesis
of~\eqref{aa2}. The integrand does not depend on $\sigma =
\omega_1+\omega_2$ and is periodic under the shift $\omega\to \omega
+1$, so we can perform the integral over $\sigma$ and obtain
\begin{equation}\label{Iinte}
 I = \int_0^\infty d \l \int_{0}^{1} d \z \int_{-\zeta}^{\zeta}
 d \rho  \int_0^{1} d \w \, \l^{\frac{p-5}{2}} \, {\cal I} \ .
\end{equation}
Since the integrand is an even function of $\rho$, we can restrict
the integral over $\rho$ to the interval from $0$ to $\zeta$. The
leading contribution in the large $E$ limit scales as $E^3$ and
comes from the region where $\rho$ is set to zero, except for the
first term of $V_s$, which has a different structure and has to be
kept exact. The subleading contribution is related to the expansion
of ${\cal I}$ at the quadratic order around $\rho=0$ and we will see
that it scales as $E$, i.e. with two power less of energy with
respect to the leading contribution. Higher orders in the $\r$
expansion are further suppressed in the large $E$ limit. So we can
further approximate the building blocks~\eqref{VsVtappr} of the
integrand and write ${\cal I} \sim {\cal I}_{l} \, {\cal I}_{s}$
with
\begin{eqnarray} \label{t9}
{\cal I}_{l}(\r) & \sim&  e^{2 \pi \l \ap s \r^2 +
  2 \pi \l \z(1-\z) \frac{\ap t}{4}} \ ,
\\ \label{t9bis}
{\cal I}_{s}(\r) &\sim& \left(4 \sin^2 \pi \w + 4 \pi^2 \l^2 \r^2
\right )^{- \frac{\ap t}{4}}  e^{4 \ap s \left(\sin^2 \pi \w +
\pi^2 \l^2 \r^2 \right )\left( e^{- 2 \pi \l \z} + e^{- 2 \pi \l (1-
\z)} \right )}  ~,
\end{eqnarray}
where we wrote explicitly only the dependence on $\r$. We then
separate the two contributions mentioned above by writing $I= I_1 +
I_2$
\begin{eqnarray}
I_1 & = & \int_0^\infty d \l \int_{0}^{1} d \z \int_{\r\sim 0}
 d \rho  \int_0^{1} d \w \, {\cal I}_{l}(\r) \,
{\cal I}_{s}(\r=0)~, \\  I_2 & = & \int_0^\infty d \l
\int_{0}^{1} d \z \int_{\r\sim 0}
 d \rho  \int_0^{1} d \w \, {\cal I}_{l}(\r)
\Big[{\cal I}_{s}(\r)-{\cal I}_{s}(\r=0)\Big]~.
\label{i1i2}
\end{eqnarray}
Let us first describe how to evaluate the leading term $I_1$,
following closely \cite{Amati:1987wq}. We can regularize the integral over $\rho$
by performing a Wick rotation on the energy of the external states
$E \to i E_e$ . Then we perform the saddle point integration
over $\r$ and, after Wick rotating back to Minkowskian energy, we
obtain
\begin{equation}
I_1 = \frac{i}{\sqrt{2\ap s}} \int_0^\infty d \l \int_{0}^{1} d \z
\int_0^{1} d \w \l^{\frac{p}{2}-3}
 e^{2 \pi \l \z(1-\z) \frac{\ap t}{4}} {\cal I}_{s}(\r=0)~.
\end{equation}
In order to deal with the integral over  $\w$,  we expand the
exponential in ${\cal I}_{s}(\r=0)$ as follows
\be
e^{4 \ap s \sin^2 \pi \l \w\left(  e^{- 2 \pi \l \z} + e^{- 2
\pi
    \l (1- \z)} \right )  } =
 \sum_{n=0}^\infty  \sum_{m=0}^\infty \frac{1}{n!m!}
 \left ( 4 \ap s \sin^2 \pi \w \right )^{n+m} e^{- 2 n \pi \l
  \z} e^{- 2 m \pi \l (1- \z)} ~,
\ee
and perform the integral by setting $x = \pi
\w$ and using
\be
B(a,b) = 2 \int_0^{\pi/2} dx \sin^{2a-1} x \cos^{2b-1} x ~. \ee
The integral over  $\l$ can then be evaluated using the integral
representation of the gamma function and the result reads
\be
I_1 = i\frac{4^{- \frac{\ap t}{4}}}{\sqrt{2 \ap s}}
\sum_{n,m=0}^\infty
\int_0^1 d \z \frac{1}{n!m!} \frac{(4\ap s)^{n+m}
\Gamma(\frac{p-4}{2}) B\left (\frac{1}{2} + n + m - \frac{\ap t}{4},
\frac{1}{2} \right )} {\pi^{\frac{p-4}{2}+1} \left [ 2n \z + 2m(1 -
\z)  - \z(1-\z) \frac{\ap
      t}{2} \right ]^{\frac{p-4}{2}}} \ .
\label{2rg9}
\ee
In this equation $\z$ plays the role of a Feynman parameter for a
diagram with two propagators and an integrated momentum over $8-p$
dimensions
\be
\int_0^1 d \z \frac{(2 \pi \ap)^{\frac{p-8}{2}}
\Gamma(\frac{p-4}{2})} {\left [ 2n \z + 2m (1 - \z) - \z(1-\z)
\frac{\ap t}{2} \right ]^{\frac{p-4}{2}}} = \int \frac{d^{8-p} {\bf
k}}{(2 \pi)^{8-p}} \ \frac{1}{\left [ 2n + \frac{\ap}{2} {\bf k}^2
\right ]\left [ 2m +
    \frac{\ap}{2} ({\bf k}-{\bf q})^2 \right ]} \ ,
\ee
where ${\bf q}$ is a ($8-p)$-dimensional vector whose norm is ${\bf
  q}^2=-t$. In terms of this momentum integral, the contribution $I_1$
  becomes
\be
I_1 = i \frac{\pi^{1-\frac{p}{2}} 4^{-\frac{\ap}{4}t}}
{\sqrt{2 \ap s}(2\pi\ap)^{\frac{p}{2}-4}}
\sum_{n,m=0}^\infty  \frac{(4\ap s)^{n+m}}{n!m!} \int \frac{d^{8-p}
{\bf k}}{(2 \pi)^{8-p}} \frac{ B\left (\frac{1}{2} + n + m -
\frac{\ap t}{4}, \frac{1}{2} \right )} {\left [ 2n + \frac{\ap}{2}
{\bf k}^2 \right ] \left [ 2m + \frac{\ap}{2} ({\bf k}-{\bf q})^2
\right ]} \ . \label{ds} \ee
We can now transform the sums over $n$ and $m$ into two
integrals over the complex plane by using
\be
\sum_{m=0}^\infty \frac{1}{m!} \frac{f(m)s^m}{m + t} = -
\int_{{\cal C}} \frac{dz_m}{2 \pi i} \frac{f(z_m)s^z_m}{z_m + t}
e^{- i \pi z_m}
\Gamma(-z_m) \ ,
\ee
and a similar expression for the sum over $n$. The contour ${\cal C}$ runs
anti-clockwise around the real positive axis in the complex plane
(clearly this identity holds if $f$ does not have any pole on the
real positive axis). Deforming the integration over the two
complex variables $z_n,\; z_m$, we pick the contributions of the
poles on the negative
real axis
\begin{eqnarray}
I_1 &\sim&
i \frac{\pi^{1-\frac{p}{2}} \left(e^{-i\pi}\ap
s\right)^{\frac{\ap}{4} t}} {\sqrt{2 \ap
s}(2\pi\ap)^{\frac{p}{2}-4}} \int  \frac{d^{8-p} {\bf k}}{(2
\pi)^{8-p}} \left\{- \Gamma\left [\frac{1}{2} - \frac{\ap}{4}t
\right ]
\frac{(4e^{-i\pi}\ap s)^{-\frac{1}{2}}} {[\ap ({\bf k} - {\bf q})^2]
[\frac{1}{2} - \frac{\ap}{4} (t + {\bf k}^2) ]}
 \nonumber \right.\\  \label{cut} &+&
\Gamma\left [\frac{\ap}{4} {\bf k}^2 \right ] \Gamma\left
[\frac{1}{2} - \frac{\ap}{4} (t + {\bf k}^2) \right]
\frac{
(4e^{-i\pi}\ap s)^{-\frac{1}{2}}} {\ap ({\bf k}^2  + ({\bf k} - {\bf
q})^2 + t)-2 }
\\  \nonumber  &+& \left.
\frac{1}{4} \Gamma\left [\frac{\ap}{4} {\bf k}^2 \right ]
\Gamma\left [\frac{\ap}{4} ({\bf q} - {\bf k})^2 \right ] B \left
(\frac{1}{2} + \frac{\ap }{2} (-{\bf k}^2  +  {\bf k} \cdot {\bf
q}), \frac{1}{2} \right )  (4 e^{-i\pi}\ap s)^{\frac{\ap}{2} (-{\bf
k}^2  + {\bf k} \cdot {\bf q})}\right\} \ ,
\end{eqnarray}
where we included only the first pole in the Euler beta function in~\eqref{ds},
since the remaining poles (in $z_n+z_m+1/2 -\ap t/4=-k$ with $k\geq
1$) yield contributions that are finite or small in the high-energy
limit. Each line in this equation contains some non-physical poles
when ${\bf k}$ takes particular values. For instance, the first term
of the integral is divergent when $2 -{\ap} (t + {\bf k}^2)=0$, but
this singularity is cancelled by a similar pole in the second line
of~\eqref{cut}. Similarly the first pole of the beta function in the last
line is cancelled by the other pole of the second term (while the
remaining poles of the beta function would be cancelled by the terms we
neglected in~\eqref{cut}). By including the prefactor in~\eqref{aa2}
and using the duplication formula $\Gamma (x) \Gamma (x + \frac{1}{2})
= \sqrt{\pi} \, 2^{1-2x} \Gamma (2x)$, we can finally rewrite the last term
of~\eqref{cut} as in~\eqref{2rg4}.

The Euler beta function in the last line of~\eqref{cut} can be written in terms of the
expectation value of two tachyon-like string vertices \cite{Amati:1987wq}
\begin{equation}\label{efft}
 \langle 0 |  \prod_{i=1}^2 \int\limits_0^{2\pi} \frac{d \s_i}{2 \pi}
: e^{i {\bf k}_i \hat{X}(\s_i)} :  | 0 \rangle = \frac{2^{\ap {\bf
k}_1 {\bf k}_2}}{\pi} B \left( \frac{1}{2}+\frac{\ap}{2} {\bf k}_1
{\bf k}_2,\frac{1}{2} \right)~,
\end{equation}
where $\hat{X}(z)$ has the standard mode expansion, but without the
zero-modes describing the centre of mass position and momentum
\begin{equation}\label{Xex}
\hat{X}(\s_i) = i \sqrt{\frac{\ap}{2}} \sum_{n\not=0}
\left(\frac{\a_n}{n} e^{i n \s}+
\frac{\bar{\a}_n}{n} e^{-i n \s}\right) \ ,
\end{equation}
which yields the OPE $\hat{X}(z) \hat{X}(w)\sim -\ap \ln|1-w/z|$.  At this point
one can simply follow the same steps as in Section~\ref{eikonal} and
use~\eqref{efft} with ${\bf k}_1 ={\bf k}$ and ${\bf k}_2 ={\bf q- k}$
in order to show that ${\cal A}^{(3)}_2$, which follows from the last
line of~\eqref{cut}, can be written as the convolution in
Eq.~\eqref{ah40}.

Let us briefly comment on the leading string corrections at small
$\ap t$ (or at large impact parameters $b$), which are
contained in the last line of~\eqref{cut}. They are obtained by
first expanding for small
$\ap t$ the r.h.s. of~\eqref{efft}, or equivalently the vertex $V_2$
in~\eqref{2rv},
\be
\label{v2expa} V_2({\bf k},{\bf q-k}) \sim 1 + \frac{\pi^2}{6}
\left(\frac{\ap}{2} {\bf k}\cdot ({\bf q-k})\right)^2 +\ldots ~,
\ee
and then by using this approximate form in~\eqref{2rg4}. One can then see
explicitly that also the contribution of these first string corrections
takes the form of a convolution involving the second derivatives of ${\cal A}_1$
as in~\eqref{secder}. After rewriting ${\cal A}_2^{(3)}$ in impact-parameter
space, we can compare this result with the
expansion of~\eqref{tash} for small $\ap Q_\perp$ and small $\ap Q_\parallel$
finding agreement at the
quadratic order. The result in~\eqref{tash} shows how these string
corrections are properly incorporated in the eikonal operator once
diagrams with an arbitrary number of boundaries are resummed.

We now analyse the subleading contribution $I_2$. For simplicity we shall
focus on the term that is non-vanishing in the field theory limit,
that is the first term in the series expansion of the exponential in the definition
of ${\cal I}_s$ in~\eqref{t9bis}. The other
terms can be evaluated in a similar way.

With the change of
variable $\sin^2\pi\w \to x$, one can see that the integral over $\w$ yields a
hypergeometric function
\begin{eqnarray}\label{f2o}
\int_0^1 d\w \left(\sin^2 \pi \w + y^2 \right )^{c}
= y^{2 c}~ {}_2 F_1(-c,1/2;1;-y^{-2}) \nonumber \\
= \frac{1}{\sqrt{\pi}} \frac{\Gamma\left(\frac{1}{2}+c\right)}{\Gamma(1+c)} +
\frac{y^{1+ 2 c}}{\sqrt{\pi}}
\frac{\Gamma\left(-\frac{1}{2}-c\right)}{\Gamma(-c)} +\ldots~,
\end{eqnarray}
where in the second line we wrote the first two terms in the
small-$y$ expansion. The first term in \eqref{f2o} is cancelled by the
contribution of ${\cal I}_s(0)$ in the definition \eqref{i1i2} and thus we obtain
\begin{equation}\label{ri}
I_2 \sim \frac{\Gamma\left(-\frac{1}{2}+\frac{\ap}{4}
  t\right)}{2\sqrt{\pi} \Gamma\left(\frac{\ap}{4} t\right)}
\int\limits_0^\infty \frac{d \l}{\l^{\frac{5-p}{2}}}
\int\limits_{0}^{1} d \z \int\limits_{\r\sim 0} d \rho
\, \left( 2 \pi \l \r \right)^{1-\frac{\ap}{2} t} e^{2 \pi \l \ap \left(s
  +\frac{t}{4}\right)\r^2 + 2 \pi \l \z(1-\z) \frac{\ap t}{4}} ~.
\end{equation}
The expansion in the second line of~\eqref{f2o} is appropriate for the
regime of very high energies at large, but fixed, values of the impact
parameter; in this way $\rho\lambda$, which is of the order of $b/(\ap
E)$, is small. If we keep increasing $b$ and keep $E$ fixed, we reach
a point where $y$ in~\eqref{f2o} is of order one and so it is more
appropriate to expand the result~\eqref{f2o} for small $c=-\ap t/4$ at
arbitrary values of $y$. By using for instance Appendix~B.1
of~\cite{Huber:2007dx}, we obtain
\begin{equation}\label{smace}
y^{2 c}~ {}_2 F_1(-c,1/2;1;-y^{-2}) =
1 - 2 c \ln2 + 2 c \pi\l\r + O(c^2)~,
\end{equation}
where in this case we used the full expression for $y=\sinh(\pi\l\r)$,
since $\l\r$ is not necessarily small. The $\r$-independent terms
coincide with the small $c$ expansion of the first term in the second
line of~\eqref{f2o} and so they have been already included in
$I_1$. The term proportional to $\pi\l\r$ in~\eqref{smace} agrees
with the small $c$ expansion of the second term in~\eqref{f2o}
(remember that in that equation $y$ is approximated with
$\pi\l\r$). This shows that the first order in $t$ of Eq.~\eqref{ri}
captures correctly the contribution of $I_2$ also at very large values
of the impact parameter $b\geq \ap E$.

At high energies, the integral over $\rho$ in Eq. \eqref{ri}
is dominated by the saddle point $\r\sim 0$ and after a Wick rotation $E\to iE_e$
we obtain the following asymptotic behaviour for $I_2$ in the
$s\gg t$ limit
\begin{equation}\label{eq93}
I_2 \sim  -
(\ap s)^{-1 +\frac{\ap}{4} t} \,  e^{- i \pi \frac{\ap}{4} t}  \, \frac{\Gamma\left(1-\frac{\ap}{4} t\right)
\Gamma\left(-\frac{1}{2}+\frac{\ap}{4}
  t\right)}{2 \sqrt{\pi} \Gamma\left(\frac{\ap}{4} t\right)}
\int\limits_0^\infty \frac{d \l}{\l^{\frac{5-p}{2}}}
\int\limits_{0}^{1} d \z \left[ 2\pi \l e^{-2\pi\l
\z(1-\z)} \right]^{-\frac{\ap}{4} t} ~.
\end{equation}
The higher powers of $\r$ in the expansion of the hypergeometric function in~\eqref{f2o},
which we have neglected, would yield contributions that are suppressed by further
powers of $1/E$. This is a consequence of
the relation between the
expansion in powers of $\rho$ and the asymptotic dependence on the energy
mentioned after Eq.~\eqref{Iinte}. The integrals over
$\z$ and $\l$ are straightforward and yield
\begin{equation}\label{I2st}
I_2 \sim (\ap s)^{-1 +\frac{\ap}{4} t} \,
\frac{\Gamma\left(1-\frac{\ap}{4} t\right) \Gamma\left(-\frac{1}{2}+\frac{\ap}{4}
  t\right)}{{2^{\frac{p-1}{2}}
\pi^{\frac{p}{2}-1}} \Gamma\left(1+\frac{\ap}{4} t\right)} \frac{
B\left(\frac{5-p}{2}+\frac{\ap}{4} t,\frac{5-p}{2}+\frac{\ap}{4} t\right)}
{\left(-\frac{\ap}{4} t\right)^{\frac{p-5}{2}-\frac{\ap}{4}t}}
\Gamma\left(\frac{p-3}{2}-\frac{\ap}{4} t\right)~.
\end{equation}
From this equation, including the prefactor in~(\ref{aa2}) and sending $\alpha' \rightarrow 0$,
one can derive Eq.~\eqref{A22}.

\sect{Dilaton scattering in field theory.}
\label{appC}

In this Appendix we provide some details about the field theory
diagrams that contribute to the dilaton scattering from a stack of
Dp-branes. As in the rest of this paper, we are interested in the
high-energy limit of the full scattering amplitude and so we can
focus on a subclass of diagrams. At the first order in
$(R_p/b)^{7-p}$ there is just a single diagram with a graviton
exchange between the probe dilaton and the Dp-branes. It is
straightforward to check that the Feynman rules summarized in
Table~\ref{tabfey} yield the high-energy result of Eq.~\eqref{T1ft}.

\begin{table}[h]
\begin{center}
\begin{tabular}{cc}
\parbox{4cm}{
\scalebox{.5}{\includegraphics{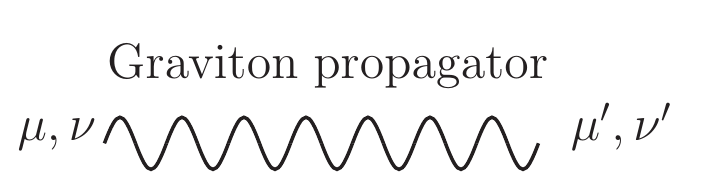}}}
 & $ \left(\eta^{\mu\mu'}\eta^{\nu\nu'} +
  \eta^{\mu\nu'}\eta^{\nu\mu'}- \frac{1}{4} \eta^{\mu\nu}
  \eta^{\mu'\nu'} \right) \frac{-i}{k^2}$
\\
\parbox{4cm}{
\scalebox{.5}{\includegraphics{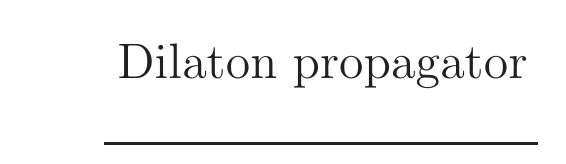}}}
 & $ \frac{-i}{k^2}$
\\
\parbox{4cm}{
\scalebox{.5}{\includegraphics{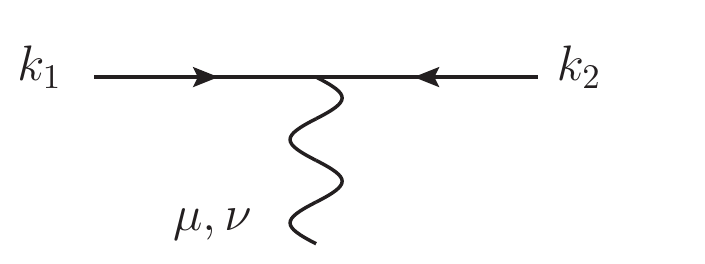}}}
 &
$ -2i \kappa \left[k_{1 (\mu} k_{2 \nu)} -\frac{1}{2} k_{1 \rho} k_2^\rho
 \eta_{\mu\nu} \right]$
\\
\parbox{4cm}{
\scalebox{.5}{\includegraphics{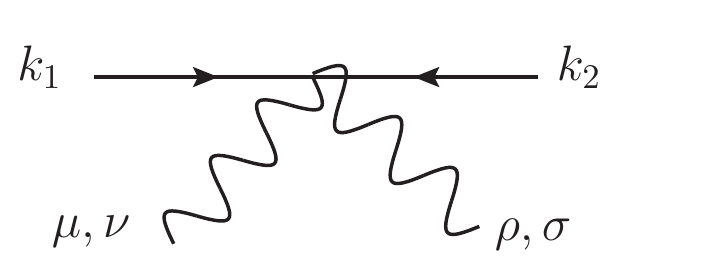}}}
 &
$ i(2\kappa)^2 \left[2 k_{1 (\mu} k_{2 (\rho} \eta_{\nu)\sigma)}
- \frac{1}{2}  (k_{1 (\mu} k_{2 \nu)} \eta_{\rho\sigma} +
k_{1 (\rho} k_{2 \sigma)} \eta_{\mu\nu}) \right] + \ldots$
\\
\parbox{4cm}{
\scalebox{.5}{\includegraphics{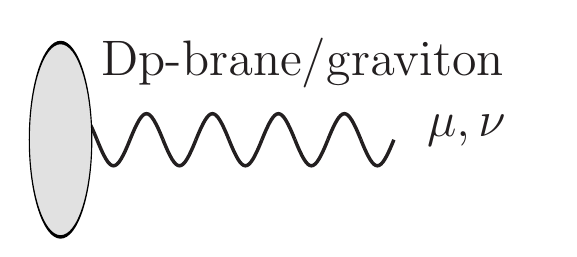}}}
 &
$-i T_p \eta_{\mu\nu}^\parallel = -i \frac{ 2 R_{p}^{7-p}
  \pi^{\frac{9-p}{2}}}{\kappa \Gamma (\frac{7-p}{2})}
\eta_{\mu\nu}^\parallel $
\end{tabular}

\caption{{\it Here we collect the Feynman rules necessary for the
  derivation of the diagrams in Figure~\ref{feyn1}. For the graviton
  propagator we use the De Donder gauge. As usual, the round
  parentheses imply a symmetrization of the enclosed indices: for
  instance we have $k_{1 (\mu} k_{2 \nu)} = (k_{1 \mu} k_{2 \nu} +
  k_{1 \nu} k_{2 \mu})/2$. The symbol $\eta_{\mu\nu}^\parallel$ means
  that the Lorentz indices are restricted to the $p+1$ directions
  along the brane world-volume. The dots in the quartic vertex stands
  for terms that are proportional to $k_1 \cdot k_2$ which cannot give
  contributions growing with $E$ in the diagrams we are interested
  in.}
\label{tabfey}}
\end{center}
\end{table}

\begin{figure}[h]
\begin{center}
\scalebox{.5}{\includegraphics{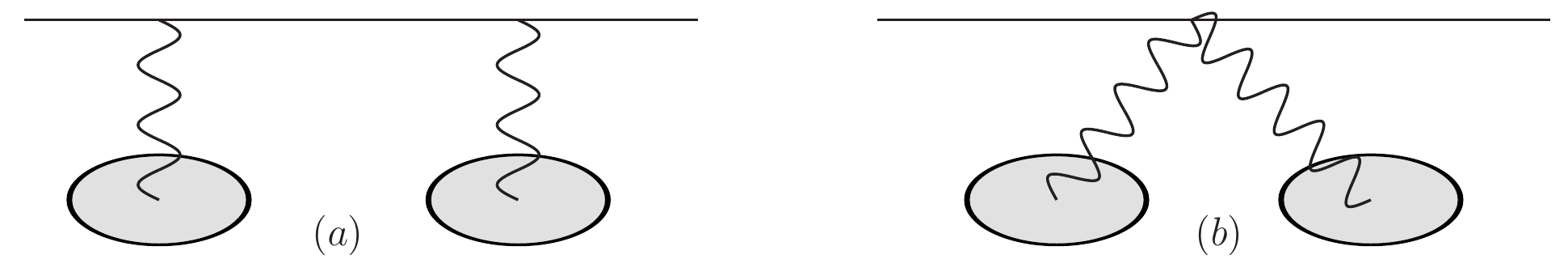}}%
\end{center}
\caption{\label{feyn1} {\it Diagram (a) contributes both to the
leading
  ($\sim E^2$) and to the subleading ($\sim E$) terms in the
  S-matrix. Diagram (b) yields contributions scaling at most as
  $E$.}}
\end{figure}

\begin{figure}[h]
\begin{center}
\scalebox{.5}{\includegraphics{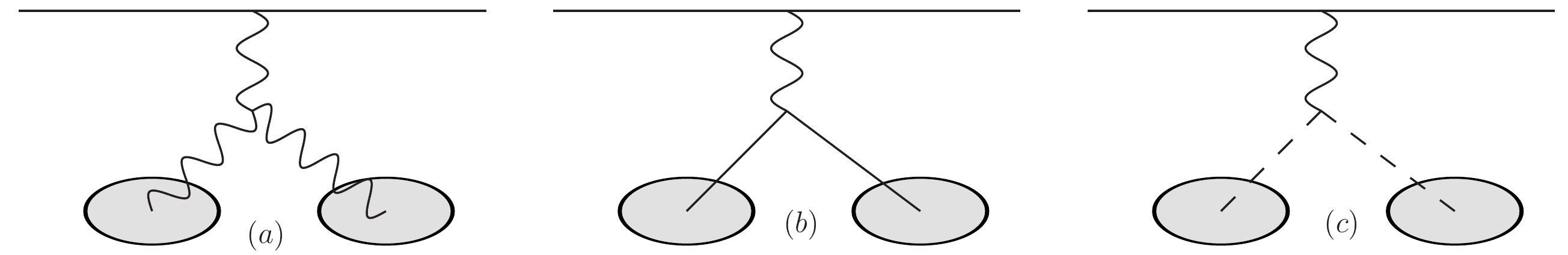}}%
\end{center}
\caption{\label{feyn2} {\it The contributions of these diagrams to
the
  S-matrix scale at most as $E$. The third diagram involves the
  exchange of a RR-field represented by the dashed lines. At
  high energies they combine and cancel exactly the leading term of
  diagram~\ref{feyn1}(b).}}
\end{figure}

We can use the same Feynman rules to compute the diagrams in
Figure~\ref{feyn1}. They scale as $(R_p/b)^{2(7-p)}$ and should
therefore contribute both to the exponentiation of the leading term,
see Eq.~\eqref{ftlim3}, and to the subleading term, see
Eq.~\eqref{A22}. We also need to consider a second class of
diagrams, depicted in Figure~\ref{feyn2}, which are of the same
order in $R_p/b$ and involve the exchange of gravitons, dilatons and
RR-fields. While diagram \ref{feyn1}(a) gives both leading $(\sim
E^2)$ and sub-leading $(\sim E)$ contributions to the S-matrix,
diagram \ref{feyn1}(b) and the three diagrams in Fig. \ref{feyn2}
give only sub-leading contributions. It is interesting to note that
diagram~\ref{feyn1}(b) cancels, in the high energy limit and in the
De Donder gauge, the diagrams in Fig.~\ref{feyn2}. As a result, both
the leading and the sub-leading contributions to the S-matrix come
from diagram \ref{feyn1}(a) alone.

Let us start our analysis from diagram~\eqref{feyn1}(a); by using
the Feynman rules we have
\begin{equation}\label{d3a}
\mbox{\ref{feyn1}(a)} =
i {  ( \kappa T_p)^2}  (2 \pi)^{p+1} \delta^{(p+1)} (p_1 + p_2)
\left[\int \frac{d^{9-p} k}{(2 \pi)^{9-p}} \frac{4E^4}{
  (p_1 -k)^{2} k^{2} (p_2 +k)^{2}} \right] ~,
\end{equation}
where $p_1,\;p_2$ are the momenta of the two external states and $k$
is the momentum flowing in the dilaton propagator. Notice that the
combinations $(p_1 -k)$ and $(p_2 +k)$ are non-trivial only in the
$(9-p)$ transverse directions, while $k$ has also a time component
and so $k^2 = -E^2 + k^2_\perp$. By performing the same Wick
rotation ($E = i E_e$) used in the string computation, we get
standard Euclidean propagators and can introduce Schwinger
parameters for evaluating~\eqref{d3a}. By focusing on the square
parenthesis we have
\begin{equation}\label{d3ai}
[\ldots] = \frac{4E_e^4}{ (4 \pi)^{\frac{9-p}{2}}}  \int_{0}^{\infty} dT
\,\,T^{- \frac{9-p}{2} +2 } \int_{0}^{1} d x \int_{0}^{x}   dy
\,\,{\rm e}^{-E_e^2 T y^2} \,{\rm e}^{-|t| T (1-x) (x-y) }~.
\end{equation}
In the high-energy limit, this integral is dominated by the saddle
point $y\sim 0$. This is the same pattern seen in the string
computation of Appendix~\ref{appA} and indeed $y$ plays the same
role as $\rho$. In order to derive the leading term, one can neglect
$y$ in the second exponential and treat the integral over $y$ in the
Gaussian approximation. By inserting the result in~\eqref{d3a} and
using~\eqref{bseries} we obtain the full result~\eqref{ftlim3} for
the dilaton amplitude at the next-to-leading order in the energy.

The subleading contribution is obtained by expanding to linear order
in $y$ the second exponential in~\eqref{d3ai}. Again by focusing on
the saddle point $y~\sim 0$ all integrals can be performed and we
obtain the contribution from diagram \ref{feyn1}(a) to ${\cal
  A}^{(2)}_2$
\begin{equation}\label{contr3a}
R_{p}^{2(7-p)} E^2 \pi^{\frac{9-p}{2}}  2^{p-6}  |t|^{\frac{5-p}{2}}
\frac{\Gamma ( \frac{p-5}{2})}{\Gamma (6-p)} ~,
\end{equation}
which agrees with the amplitude ${\cal A}^{(2)}_2$ of~\eqref{A22}.

Diagram \ref{feyn1}(b) yields
\begin{eqnarray}\label{d3b}
\mbox{\ref{feyn1}(b)} & =&
\frac{i}{2} {  ( \kappa T_p)^2}  (2 \pi)^{p+1} \delta^{(p+1)} (p_1 + p_2)
\left[\int \frac{d^{9-p} q}{(2 \pi)^{9-p}}
\frac{E^2 (7-p)}{q^2 (q-(p_1+p_2))^2} \right] +\ldots
\\ \nonumber &=&  \left(i (2 \pi)^{p+1} \delta^{(p+1)} (p_1 + p_2)\right)
 R_{p}^{2(7-p)} E^2 \pi^{\frac{9-p}{2}} 2^{p-6} |t|^{\frac{5-p}{2}}
\frac{\Gamma (\frac{p-5}{2})}{\Gamma ( 6-p) } \frac{(7-p)}{4 (6-p)}+\ldots ~,
\end{eqnarray}
where $q$ is the momentum of one of the two graviton propagators and
the dots stand for subleading terms in $E$. The leading energy
result in~\eqref{d3b} cancels the analogue contribution from the
Feynman diagrams of Figure~\ref{feyn2}. The direct evaluation of
these diagrams is somewhat involved, in particular for the
contribution containing the 3-graviton vertex. However, as an
intermediate check, one can combine the three diagrams, before
sewing the common dilaton-dilaton-graviton vertex,  and see that,
after a Fourier transformation, this reproduces the Dp-brane metric
in the Einstein frame at the second order in the $(R_p/r)^{7-p}$
expansion. After including the vertex with the external states, we
obtain, at leading order in $E$, a contribution opposite
to~\eqref{d3b}. This shows that, in the gauge we chose, all
contributions growing with the energy are captured by diagram
\ref{feyn1}(a). It would be very interesting to check whether a
similar pattern holds also at higher orders in $R_p/b$ and to see
whether the only relevant contributions at high energy are the
``half-ladder'' diagrams which are a natural generalization of
\ref{feyn1}(a).

\sect{Computation of $c_0$, $c_x$ and $c_y$}
\label{appB}

Let us start by showing that $c_x$ in~\eqref{cxy0} is vanishing at
order $\left(\frac{R_{p}}{b}\right)^{7-p}$. As mentioned in the main
text, the first term in the square parenthesis in~\eqref{ax} is
clearly of higher order, since $\ln\sqrt{\a}\sim O (R_p/r)^{7-p}$. The
second term is a total derivative and its contribution
\begin{equation}
\partial_u \ln\sqrt{\alpha}\,\Big|^{\infty}_0
= \left[\frac{C}{\b} \partial_r  \ln\sqrt{\alpha}
\;\right]^{\infty}_{r_*} =0~,
\end{equation}
where we first used~\eqref{C} to change the derivative with respect to $u$
into a derivative with respect to $r$ and then used $C(r_*)=0$.

A similar computation holds also for $c_y$. The combination $r
\sin\bar\theta= b +O\left((R/r)^{7-p}\right)$ and so again the first
term on the r.h.s. of~\eqref{ay} can be neglected at the order
$\left(\frac{R_{p}}{b}\right)^{7-p}$. At leading order we have
\begin{equation}\label{cyb2}
c_y \sim 2 \partial_u\ln[\sqrt{\b} r \sin\bar\theta(u)] \Big|^\infty_0 =
\left[\frac{C}{\b} \partial_r \ln[\sqrt{\b}r\sin\theta(r)]
\right]^\infty_{r_*} ~.
\end{equation}
The function $C(r)$ vanishes at the turning point $r_*$, so the only
possibility for obtaining a non-zero result is to act with the
derivative on the $\sin\theta$ term. From~\eqref{eqtraj} we see
that this yields a factor of $1/C$ and the lower extremum
of~\eqref{cyb2} yields the non-vanishing contribution
\be
2 \left.\frac{b\cos\theta}{r^2\b\sin\theta}\right|_{r=r_*} =
2 \frac{b\cos\left(\frac{\Theta_p+\pi}{2}\right)}
{r_*^2\b(r_*)\sin\left(\frac{\Theta_p+\pi}{2}\right)}
\sim -\frac{\Theta_p}{b}~,
\ee
where in the first step we used the relation between $\theta(r_*)$ and
$\Theta_p$ in Eq.~\eqref{8a} and then we kept only the terms of
order $\left(\frac{R_p}{b}\right)^{7-p}$. Using~\eqref{leadnext}, we
obtain the result~\eqref{cyre}.

Let us finally consider the coefficient $c_0$. From Eq.~(\ref{A}) we
have
\eq c_0 = \int_{-\infty}^{+\infty}
\frac{\partial_u^2 \sqrt{\beta r^2 - b^2 \alpha}}{\sqrt{\beta r^2 -
    b^2 \alpha}}  du \ ,
\eqx
which we can rewrite as
\eq
\label{c0}
c_0 = 2 \int_{r_*}^{\infty} \frac{dr}{ \sqrt{\beta r^2 - b^2 \alpha}}
\partial_r \partial_u \sqrt{\beta r^2 - b^2 \alpha} \ .
\eqx
We then compute
\eq
\partial_u \sqrt{\beta r^2 - b^2 \alpha} =
\frac{\partial r}{\partial u }\partial_r \sqrt{\beta r^2 - b^2 \alpha}=
\alpha^{-1/2}\left( 1 + \frac{r \beta'}{2\beta} - \frac{b^2}{2r}
\frac{\alpha'}{\beta} \right) \ , \label{du}
\eqx
where we have used Eq. $(\ref{C})$ and a prime denotes differentiation
with respect to $r$. Since we are interested only in the lowest order
terms that behave as $\left(\frac{R_p}{b}\right)^{7-p}$, we can expand
$\alpha$ and $\beta$ at large $r$ as follows
\eq
\alpha \sim 1 + C_{\alpha}  \left(\frac{R_p}{r}\right) ^{7-p} \ , \hspace{1cm}
\beta \sim 1 +  C_{\beta}
\left(\frac{R_p}{r}\right) ^{7-p} \ .
\label{ab}
\eqx
We keep the coefficients $C_{\alpha}$ and $C_{\beta}$ arbitrary at this stage, but we
will see that the result depends only on the difference
$C_{\beta}- C_{\alpha}$, which is equal to one both in the string and
in the Einstein frame. Substituting Eqs.~$(\ref{ab})$ in~$(\ref{du})$ and taking
the derivative with respect to $r$ we get
\begin{eqnarray}
&&\partial_r \left[  \frac{1}{2} \left(\frac{R_p}{b}\right)^{7-p}
    \left(  - C_{\alpha} + (p-7) C_{\beta} - \frac{b^2}{r^2} (p-7)
    C_{\alpha} \right) \right] \nonumber \\
&&= \frac{p-7}{2r} \left(\frac{R_p}{r}\right)^{7-p}  \left[ -
    C_{\alpha} + C_{\beta} (p-7) - \frac{b^2}{r^2} C_{\alpha} (p-9)
    \right] \ .
\label{for}
\end{eqnarray}
We can now insert this expression in Eq.~$(\ref{c0})$
and make the approximations, valid at the lowest order that we are considering,
$r^{*} \sim b$ in the lower extremum of integration
and $\alpha$, $\beta \sim 1$ in the square root
in the denominator. Using the integral
\begin{eqnarray}
\int_{b}^{\infty}  dr \, \frac{ r^{1 - 2 \gamma}}{\sqrt{r^2- b^2}} =
\frac{  \sqrt{\pi} \,\,b^{1 - 2 \gamma} }{2}
\frac{ \Gamma (\gamma - \frac{1}{2} ) }{ \Gamma ( \gamma)} \ ,
\label{form}
\end{eqnarray}
we obtain
\begin{eqnarray}
c_0 = \frac{ \sqrt{\pi}}{b} (7-p)
\left(\frac{R_p}{b}\right)^{7-p}
\frac{\Gamma ( \frac{ 8-p}{2} )  }{\Gamma ( \frac{7-p}{2})}
(C_{\beta} - C_{\alpha}) \ .
\label{cob}
\end{eqnarray}
Since $C_{\beta} - C_{\alpha}=1$, the previous result reproduces  Eq.~(\ref{c00}).

\providecommand{\href}[2]{#2}\begingroup\raggedright\endgroup

\end{document}